\documentclass[apj]{emulateapj}
\usepackage{apjfonts,graphicx,here}
\begin{document}
\title{Bandpass Dependence of X-ray Temperatures in Galaxy Clusters}
\author{Kenneth W. Cavagnolo\altaffilmark{1,2}, Megan
Donahue\altaffilmark{1}, G. Mark Voit\altaffilmark{1}, and Ming Sun\altaffilmark{1}}
\altaffiltext{1}{Michigan State University, Department of Physics and Astronomy, BPS Building, East Lansing, MI 48824}
\altaffiltext{2}{cavagnolo@pa.msu.edu}
\shorttitle{X-ray Band Dependent Temperature}
\shortauthors{K. W. Cavagnolo et al.}

\journalinfo{To appear in the Astrophysical Journal}
\accepted{March 26, 2008}
\submitted{ }
\begin{abstract}

We explore the band dependence of the inferred X-ray temperature of
the intracluster medium (ICM) for 192 well-observed galaxy clusters
selected from the {\it Chandra} Data Archive. If
the hot ICM is nearly isothermal in the projected region
of interest, the X-ray temperature inferred from a broad-band
(0.7-7.0 keV) spectrum should be identical to the X-ray temperature
inferred from a hard-band (2.0-7.0 keV) spectrum. However, if 
unresolved cool lumps of gas are contributing soft X-ray emission, the
temperature of a best-fit single-component thermal model will be
cooler for the broad-band spectrum than for the hard-band spectrum. Using
this difference as a diagnostic, the ratio of best-fitting hard-band
and broad-band temperatures may indicate the presence of cooler gas even
when the X-ray spectrum itself may not have sufficient signal-to-noise
to resolve multiple temperature components. To test this possible
diagnostic, we extract X-ray spectra from core-excised annular regions
for each cluster in our archival sample. We compare the X-ray
temperatures inferred from single-temperature fits when the energy
range of the fit is 0.7-7.0 keV (broad) and when the energy range is
2.0/(1+$z$)-7.0 keV (hard). We find that the hard-band temperature is
significantly higher, on average, than the broad-band
temperature. Upon further exploration, we find this temperature ratio
is enhanced preferentially for clusters which are known merging
systems. In addition, cool-core clusters tend to have best-fit hard-band
temperatures that are in closer agreement with their best-fit broad-band
temperatures. We show, using simulated spectra, that this diagnostic is
sensitive to secondary cool components ($T_X = 0.5-3.0$ keV) with
emission measures $\geq 10-30\%$ of the primary hot component.
\end{abstract}


\section{Introduction}\label{sec:intro}

The normalization, shape, and evolution of the cluster mass function
are useful for measuring cosmological parameters
(e.g. \citealt{1989ApJ...341L..71E, 1998ApJ...508..483W,
2001ApJ...553..545H,  2004PhRvD..70l3008W}). In particular, the
evolution of large scale structure formation provides a complementary
and distinct constraint on cosmological parameters to those tests
which constrain them geometrically, such as supernovae
\citep{1998AJ....116.1009R, 2007ApJ...659...98R} and baryon acoustic
oscillations \citep{2005ApJ...633..560E}. However, clusters are a
useful cosmological tool only if we can infer cluster masses from
observable properties such as X-ray luminosity, X-ray temperature,
lensing shear, optical luminosity, or galaxy velocity
dispersion. Empirically, the correlation of mass to these observable
properties is well-established (see \cite{2005RvMP...77..207V} for a review). But, there
is non-negligible scatter in mass-observable scaling relations which must be
accounted for if clusters are to serve as high-precision mass proxies
necessary for using clusters to study cosmological parameters such as
the dark energy equation of state. However, if we could identify a
``2nd parameter" -- possibly reflecting the degree of relaxation in
the cluster -- we could improve the utility of clusters as
cosmological probes by parameterizing and reducing the scatter in
mass-observable scaling relations.

Toward this end, we desire to quantify the dynamical state of a
cluster beyond simply identifying which clusters appear relaxed and
those which do not. Most clusters are likely to have a dynamical state
which is somewhere in between \citep{2006ApJ...639...64O,
2006ApJ...650..128K, VV08}. The degree to which a cluster is
virialized must first be quantified within simulations that correctly
predict the observable properties of the cluster. Then, predictions
for quantifying cluster virialization may be tested, and possibly
calibrated, with observations of an unbiased sample of clusters
(e.g. REXCESS sample of \citealt{2007A&A...469..363B}).

One study that examined how relaxation might affect the observable
properties of clusters was conducted by \citealt{2001ApJ...546..100M}
(hereafter ME01) using the ensemble of simulations by
\citealt{1997ApJ...491...38M}. ME01 found that most clusters which had
experienced a recent merger were cooler than the cluster
mass-observable scaling relations predicted. They attributed this to the
presence of cool, spectroscopically unresolved accreting subclusters
which introduce energy into the ICM and have a long timescale for
dissipation. The consequence was an under-prediction of cluster
binding masses of $15-30\%$ \citep{2001ApJ...546..100M}. It is
important to note that the simulations of \cite{1997ApJ...491...38M}
included only gravitational processes. The intervening years have
proven that radiative cooling is tremendously important in shaping the
global properties of clusters (e.g. \citealt{2004ApJ...613..811M},
\citealt{2006MNRAS.373..881P}, or
\citealt{2007ApJ...668....1N}). Therefore, the magnitude of the effect
seen by ME01 could be somewhat different if radiative processes are
included.

One empirical observational method of quantifying the degree of
cluster relaxation involves using ICM substructure and employs the power in
ratios of X-ray surface brightness moments \citep{1995ApJ...452..522B,
1996ApJ...458...27B, 2005ApJ...624..606J}. Although an excellent tool,
power ratios suffer from being aspect-dependent
\citep{2007arXiv0708.1518J, VV08}. The work of ME01 suggested a
complementary measure of substructure which does not depend on
projected perspective. In their analysis, they found hard-band
(2.0-9.0 keV) temperatures were $\sim 20\%$ hotter than broad-band
(0.5-9.0 keV) temperatures. Their interpretation was that the cooler
broad-band temperature is the result of unresolved accreting cool
subclusters which are contributing significant amounts of line
emission to the soft band ($E < 2$ keV). This effect has been studied
and confirmed by \cite{2004MNRAS.354...10M} and
\cite{2006ApJ...640..710V} using simulated {\it Chandra} and
{\it{XMM-Newton}} spectra.

ME01 suggested that this temperature skewing, and consequently the
fingerprint of mergers, could be detected utilizing the energy
resolution and soft-band sensitivity of {\it Chandra}. They proposed
selecting a large sample of clusters covering a broad dynamical range,
fitting a single-component temperature to the hard-band and
broad-band, and then checking for a net skew above unity in the
hard-band to broad-band temperature ratio. In this paper we present
the findings of just such a temperature-ratio test using {\it Chandra}
archival data. We find the hard-band  temperature exceeds the
broad-band temperature, on average, by $\sim16\%$ in multiple
flux-limited samples of X-ray clusters from the {\it Chandra}
archive. This mean excess is weaker than the $20\%$ predicted by ME01,
but is significant at the $12\sigma$ level nonetheless. Hereafter, we
refer to the hard-band to broad-band temperature ratio as
$T_{HBR}$. We also find that non-cool core systems and mergers tend to
have higher values of $T_{HBR}$. Our findings suggest that $T_{HBR}$
is an indicator of a cluster's temporal proximity to the most recent
merger event.

This paper proceeds in the following manner:
In \S\ref{sec:selection} we outline sample-selection criteria and {\it
Chandra} observations selected under these criteria. Data reduction
and handling of the X-ray background is discussed in
\S\ref{sec:data}. Spectral extraction is discussed in
\S\ref{sec:extraction}, while fitting and simulated spectra are
discussed in \S\ref{sec:specan}. Results and discussion of our
analysis are presented in \S\ref{sec:r&d}. A summary of our work is
presented in \S\ref{sec:summary}. For this work we have assumed a flat
$\Lambda$CDM Universe with cosmology $\Omega_{M} = 0.3$,
$\Omega_{\Lambda} = 0.7$, and $H_{0} = 70$ km s$^{-1}$ Mpc$^{-1}$. All
quoted uncertainties are at the 1.6$\sigma$ level (90\% confidence).

\section{Sample Selection} \label{sec:selection}

Our sample was selected from observations publicly available in the
{\it Chandra} X-ray Telescope's Data Archive (CDA). Our initial
selection pass came from the {\it{ROSAT}} Brightest Cluster Sample
\citep{1998MNRAS.301..881E}, RBC Extended Sample
\citep{2000MNRAS.318..333E}, and {\it{ROSAT}} Brightest 55 Sample
\citep{1990MNRAS.245..559E, 1998MNRAS.298..416P}. The portion of our
sample at $z \gtrsim 0.4$ can also be found in a combination of the
{\it{Einstein}} Extended Medium Sensitivity Survey
\citep{1990ApJS...72..567G}, North Ecliptic Pole Survey
\citep{2006ApJS..162..304H}, {\it{ROSAT}} Deep Cluster Survey
\citep{1995ApJ...445L..11R}, {\it{ROSAT}} Serendipitous Survey
\citep{1998ApJ...502..558V}, and Massive Cluster Survey
\citep{2001ApJ...553..668E}. We later extended our sample to include
clusters found in the REFLEX Survey \citep{2004A&A...425..367B}. Once
we had a master list of possible targets, we cross-referenced this
list with the CDA and gathered observations where a minimum of
$R_{5000}$ (defined below) is fully within the CCD field of
view.

$R_{\Delta_c}$ is defined as the radius at which the average cluster
density is $\Delta_c$ times the critical density of the Universe,
$\rho_c=3H(z)^2/8\pi G$. For our calculations of $R_{\Delta_c}$ we
adopt the relation from \cite{2002A&A...389....1A}:
\begin{eqnarray}
R_{\Delta_c} &=& 2.71 \mathrm{~Mpc~}
\beta_T^{1/2}
\Delta_{\mathrm{z}}^{-1/2}
(1+z)^{-3/2}
\left(\frac{kT_X}{10 \mathrm{~keV}}\right)^{1/2}\\
\Delta_z &=& \frac{\Delta_c \Omega_M}{18\pi^2\Omega_z} \nonumber \\
\Omega_z &=& \frac{\Omega_M (1+z)^3}{[\Omega_M
(1+z)^3]+[(1-\Omega_M-\Omega_{\Lambda})(1+z)^2]+\Omega_{\Lambda}} \nonumber
\end{eqnarray}
where $R_{\Delta_c}$ is in units of $h_{70}^{-1}$, $\Delta_c$ is
the assumed density contrast of the cluster at $R_{\Delta_c}$, and
$\beta_T$ is a numerically determined, cosmology-independent
($\lesssim \pm 20\%$) normalization for the virial relation $GM/2R =
\beta_TkT_{vir}$. We use $\beta_T = 1.05$ taken from
\cite{1996ApJ...469..494E}.

\begin{figure}
\begin{center}
\includegraphics*[width=\textwidth, trim=0mm 0mm 0mm 0mm, clip]{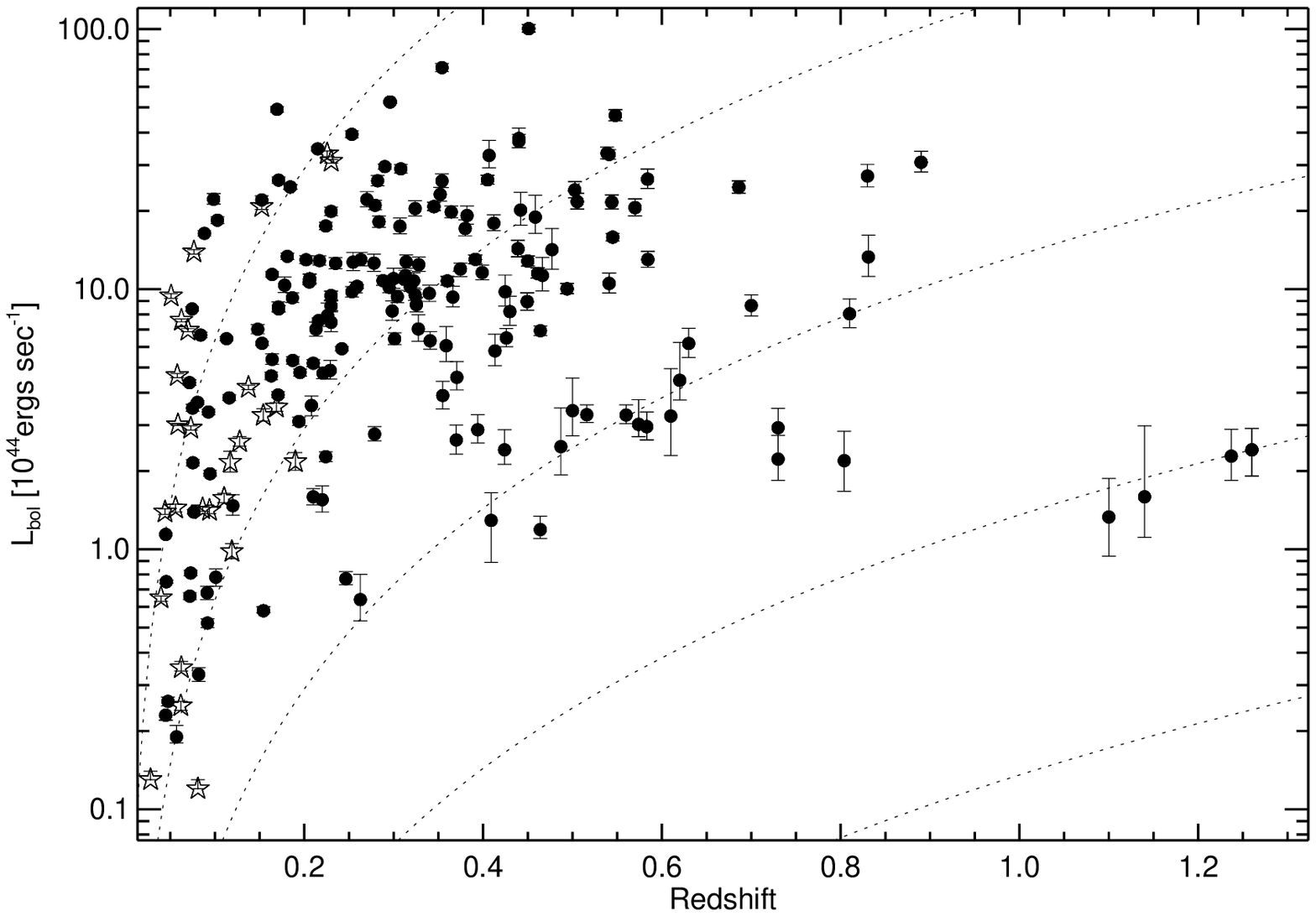}
\caption{
Bolometric luminosity ($E = 0.1-100$ keV) plotted as a function of
redshift for the 202 clusters which make up the initial
sample. $L_{bol}$ values are limited to the region of spectral
extraction, $R=R_{2500-\mathrm{CORE}}$. For clusters without
$R_{2500-\mathrm{CORE}}$ fits, $R=R_{5000-\mathrm{CORE}}$ fits were
used and are denoted in the figure by empty stars. Dotted lines
represent constant fluxes of $3.0\times10^{-15}$, $10^{-14}$,
$10^{-13}$, and $10^{-12}$ ergs sec$^{-1}$ cm$^{-2}$.
}
\label{fig:lx_z}
\end{center}
\end{figure}

The result of our CDA search was a total of 374 observations of which
we used 244 for 202 clusters. The clusters making up our sample cover
a redshift range of $z = 0.045-1.24$, a temperature range of
$T_X = 2.6-19.2 \mathrm{~keV}$, and bolometric luminosities of
$L_{bol} = 0.12-100.4\times10^{44} \mathrm{~ergs~s}^{-1}$. The
bolometric ($E = 0.1-100$ keV) luminosities for our sample clusters
plotted as a function of redshift are shown in Figure
\ref{fig:lx_z}. These $L_{bol}$ values are calculated from our
best-fit spectral models and are limited to the region of the spectral
extraction (from $R=70$ kpc to $R=R_{2500}$, or $R_{5000}$ in the
cases where no $R_{2500}$ fit was possible). Basic properties of our
sample are listed in Table \ref{tab:sample}.

For the sole purpose of defining extraction regions based on fixed
overdensities as discussed in \S\ref{sec:extraction}, fiducial
temperatures (measured with {\it ASCA}) and redshifts were taken from
the Ph.D. thesis of Don
Horner\footnote{http://asd.gsfc.nasa.gov/Donald.Horner/thesis.html}
(all redshifts confirmed with
NED\footnote{http://nedwww.ipac.caltech.edu/}). We will show later
that the {\it ASCA} temperatures are sufficiently close to the {\it
Chandra} temperatures such that $R_{\Delta_c}$ is reliably estimated
to within 20\%. Note that $R_{\Delta_c}$ is proportional to $T^{1/2}$,
so that a 20\% error in the temperature leads to only a 10\% error in
$R_{\Delta_c}$, which in turn has no detectable effect on our final
results. For clusters not listed in Horner's thesis, we used a
literature search to find previously measured temperatures. If no
published value could be located, we measured the global temperature
by recursively extracting a spectrum in the region $0.1<r<0.2
R_{500}$ fitting a temperature and recalculating $R_{500}$. This
process was repeated until three consecutive iterations produced
$R_{500}$ values which differed by $\leq 1\sigma$. This method of
temperature determination has been employed in other studies, see
\cite{2006MNRAS.tmp.1068S} and \cite{2006ApJS..162..304H} as
examples.

\section{{\it Chandra} Data}\label{sec:data}

\subsection{Reprocessing and Reduction}\label{sec:reprocessing}

All datasets were reduced utilizing the {\it Chandra} Interactive
Analysis of Observations package ({\textsc{CIAO}}) and accompanying
Calibration Database ({\textsc{CALDB}}). Using {\textsc{CIAO
v3.3.0.1}} and {\textsc{CALDB v3.2.2}}, standard data analysis was
followed for each observation to apply the most up-to-date
time-dependent gain correction and when appropriate, charge transfer
inefficiency correction \citep{2000ApJ...534L.139T}.

Point sources were identified in an exposure-corrected events file
using the adaptive wavelet tool {\textsc{wavdetect}}
\citep{2002ApJS..138..185F}. A $2\sigma$ region surrounding each point
source was automatically output by {\textsc{wavdetect}} to define an
exclusion mask.  All point sources were then visually confirmed and we
added regions for point sources which were missed by
{\textsc{wavdetect}} and deleted regions for spuriously detected
``sources''. Spurious sources are typically faint CCD features (chip
gaps and chip edges) not fully removed after dividing by the exposure
map. This process resulted in an events file (at ``level 2'') that has
been cleaned of point sources.

To check for contamination from background flares or periods of
excessively high background, light curve analysis was performed using
Maxim Markevitch's contributed {\textsc{CIAO}} script
{\textsc{lc\_clean.sl}}\footnote{http://cxc.harvard.edu/contrib/maxim/acisbg/}.
Periods with count rates $\geq 3\sigma$ and/or a factor $\geq 1.2$ of
the mean background level of the observation were removed from the
good time interval file. As prescribed by Markevitch's
cookbook\footnote{http://cxc.harvard.edu/contrib/maxim/acisbg/COOKBOOK},
ACIS front-illuminated (FI) chips were analyzed in the $0.3-12.0$ keV
range, and the $2.5-7.0$ keV energy range for the ACIS
back-illuminated (BI) chips.

When a FI and BI chip were both active during an observation, we
compared light curves from both chips to detect long duration,
soft-flares which can go undetected on the FI chips but show up on the
BI chips. While rare, this class of flare must be filtered out of the
data, as it introduces a spectral component which artificially
increases the best-fit temperature via a high energy tail. We find
evidence for a long duration soft flare in the observations of Abell
1758 \citep{2004ApJ...613..831D}, CL J2302.8+0844, and IRAS
09104+4109. These flares were handled by removing the time period of
the flare from the GTI file.

Defining the cluster ``center'' is essential for the later purpose of
excluding cool cores from our spectral analysis (see
\S\ref{sec:extraction}). To determine the cluster center, we
calculated the centroid of the flare cleaned, point-source
free level-2 events file filtered to include only photons in the
$0.7-7.0$ keV range. Before centroiding, the events file was
exposure-corrected and ``holes'' created by excluding point sources
were filled using interpolated values taken from a narrow annular region just
outside the hole (holes are not filled during spectral extraction
discussed in \S\ref{sec:extraction}). Prior to centroiding, we defined
the emission peak by heavily binning the image, finding the peak value
within a circular region extending from the peak to the chip edge
(defined by the radius $R_{max}$), reducing $R_{max}$ by 5\%,
reducing the binning by a factor of two, and finding the peak
again. This process was repeated until the image was unbinned (binning
factor of one). We then returned to an unbinned image with an aperture
centered on the emission peak with a radius $R_{max}$ and found the
centroid using {\textsc{CIAO}}'s {\textsc{dmstat}}. The centroid, ($x_c$,$y_c$),
for a distribution of $N$ good pixels with coordinates ($x_i$,$y_j$)
and values f($x_i$,$y_j$) is defined as:
\begin{eqnarray}
Q &=& \sum_{i,j=1}^N f(x_i,y_i) \\
x_c &=& \frac{\sum_{i,j=1}^N x_i \cdot f(x_i,y_i)}{Q} \nonumber \\
y_c &=& \frac{\sum_{i,j=1}^N y_i \cdot f(x_i,y_i)}{Q}. \nonumber
\end{eqnarray}

If the centroid was within 70 kpc of the emission peak, the emission
peak was selected as the center, otherwise the centroid was used
as the center. This selection was made to ensure all ``peaky'' cool
cores coincided with the cluster center, thus maximizing their
exclusion later in our analysis. All cluster centers were additionally
verified by eye.

\subsection{X-ray Background} \label{sec:background}

Because we measured a global cluster temperature, specifically looking
for a temperature ratio shift in energy bands which can be
contaminated by the high-energy particle background or the soft local
background, it was important to carefully analyze the background and
subtract it from our source spectra. Below we outline three steps
taken in handling the background: customization of blank-sky backgrounds,
re-normalization of these backgrounds for variation of hard-particle
count rates, and fitting of soft background residuals.

We used the blank-sky observations of the X-ray background from
\cite{2001ApJ...562L.153M} and supplied within the CXC
{\textsc{CALDB}}. First, we compared the flux from the diffuse soft
X-ray background of the {\it{ROSAT}} All-Sky Survey ({\it RASS})
combined bands $R12$, $R45$, and $R67$ to the 0.7-2.0 keV flux in each
extraction aperture for each observation. {\it RASS} combined bands
give fluxes for energy ranges of 0.12-0.28 keV, 0.47-1.21 keV, and
0.76-2.04 keV respectively corresponding to $R12$, $R45$, and $R67$. For
the purpose of simplifying subsequent analysis, we discarded
observations with an $R45$ flux $\geq 10\%$ of the total cluster X-ray
flux.

The appropriate blank-sky dataset for each observation was
selected from the {\textsc{CALDB}}, reprocessed exactly as the
observation was, and then reprojected using the aspect solutions provided
with each observation. For observations on the ACIS-I array, we
reprojected blank-sky backgrounds for chips I0-I3 plus chips S2 and/or
S3. For ACIS-S observations, we created blank-sky backgrounds for
the target chip, plus chips I2 and/or I3. The additional off-aimpoint
chips were included only if they were active during the observation
and had available blank-sky data sets for the observation time
period. Off-aimpoint chips were cleaned for point sources and diffuse
sources using the method outlined in \S\ref{sec:reprocessing}.

The additional off-aimpoint chips were included in data reduction
since they contain data which is farther from the cluster center and
are therefore more useful in analyzing the observation background. For
observations which did not have a matching off-aimpoint blank-sky
background, a source-free region of the active chips is
located and used for background normalization. To normalize the hard
particle component we measured fluxes for identical regions in the
blank-sky field and target field in the 9.5-12.0 keV range. The
effective area of the ACIS arrays above 9.5 keV is approximately zero,
and thus the collected photons there are exclusively from the particle
background.

\begin{figure}
\begin{center}
\includegraphics*[width=\textwidth, trim=5mm 0mm 0mm 0mm, clip]{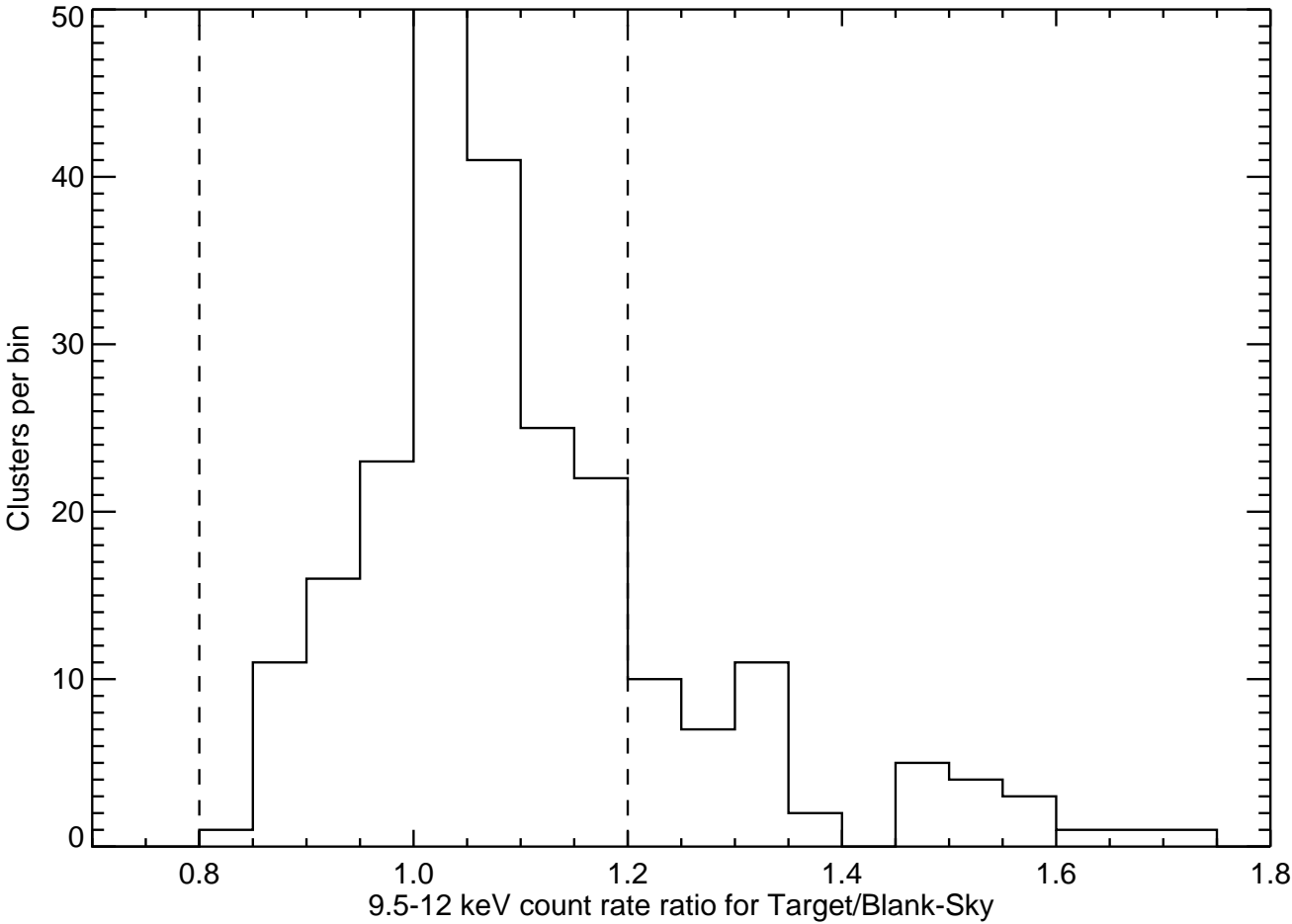}
\caption{
Ratio of target field and blank-sky field count rates in the 9.5-12.0
keV band for all 244 observations in our initial sample. Vertical
dashed lines represent $\pm 20\%$ of unity. Despite the good agreement
between the blank-sky background and observation count rates for most
observations, all backgrounds are normalized.
}
\label{fig:bgd}
\end{center}
\end{figure}

A histogram of the ratios of the 9.5-12.0 keV count rate from an
observation's off-aimpoint chip to that of the observation specific
blank-sky background are presented in Figure \ref{fig:bgd}. The
majority of the observations are in agreement to $\lesssim 20\%$ of
the blank-sky background rate, which is small enough to not affect our
analysis. Even so, we re-normalized all blank-sky backgrounds to match
the observed background.

Normalization brings the observation background and blank-sky
background into agreement for $E > 2$ keV, but even after
normalization, typically, there may exist a soft excess/deficit
associated with the spatially varying soft Galactic
background. Following the technique detailed in
\cite{2005ApJ...628..655V}, we constructed and fit soft residuals for
this component. For each observation we subtracted a spectrum of the
blank-sky field from a spectrum of the off-aimpoint field to create a
soft residual. The residual was fit with a solar abundance,
zero-redshift {\textsc{MeKaL}} model \citep{1985A&AS...62..197M,
1986A&AS...65..511M, 1992SRON, 1995ApJ...438L.115L} where the
normalization was allowed to be negative. The resulting best-fit
temperatures for all of the soft residuals identified here were
between 0.2-1.0 keV, which is in agreement with results of
\cite{2005ApJ...628..655V}. The model normalization of this background
component was then scaled to the cluster sky area. The re-scaled
component was included as a fixed background component during fitting
of a cluster's spectra.

\section{Spectral Extraction} \label{sec:extraction}

The simulated spectra calculated by ME01 were analyzed in a broad
energy band of $0.5-9.0$ keV and a hard energy band of
$2.0_{\mathrm{rest}}-9.0$ keV, but to make a reliable comparison with
{\it{Chandra}} data we used narrower energy ranges of 0.7-7.0 keV for
the broad energy band and $2.0_{\mathrm{rest}}-7.0$ keV for the hard
energy band. We excluded data below $0.7$ keV to
avoid the effective area and quantum efficiency variations of the ACIS
detectors, and excluded energies above $7.0$ keV in which diffuse source
emission is dominated by the background and where {\it{Chandra}}'s
effective area is small. We also accounted for cosmic redshift by
shifting the lower energy boundary of the hard-band from 2.0 keV to
$2.0/(1+z)$ keV (henceforth, the 2.0 keV cut is in the rest
frame).

ME01 calculated the relation between $T_{0.5-9.0}$ and $T_{2.0-9.0}$
using apertures of $R_{200}$ and $R_{500}$ in size. While it is
trivial to calculate a temperature out to $R_{200}$ or $R_{500}$ for
a simulation, such a measurement at these scales is extremely
difficult with {\it Chandra} observations (see \cite{2005ApJ...628..655V} for
a detailed example). Thus, we chose to extract spectra from regions
with radius $R_{5000}$, and $R_{2500}$ when possible. Clusters
analyzed only within $R_{5000}$ are denoted in Table \ref{tab:sample}
by a double dagger ($\ddagger$).

The cores of some clusters are dominated by gas at $\lesssim
T_{virial}/2$ which can greatly affect the global best-fit
temperature; therefore, we excised the central 70 kpc of each
aperture. These excised apertures are denoted by ``-CORE'' in the
text. Recent work by \cite{2007astro.ph..3504M} has shown excising
0.15 $R_{500}$ rather than a static 70 kpc reduces scatter in
mass-observable scaling relations. However, our smaller excised region
seems sufficient for this investigation because for cool core clusters
the average radial temperature at $r > 70$ kpc is approximately
isothermal \citep{2005ApJ...628..655V}. Indeed, we find that cool core
clusters have smaller than average $T_{HBR}$ when the 70 kpc region
has been excised (\S\ref{sec:ccncc}).

Although some clusters are not circular in projection, but rather are
elliptical or asymmetric, we found that assuming spherical symmetry
and extracting spectra from a circular annulus did not significantly
change the best-fit values. For another such example see
\cite{2005MNRAS.359.1481B}.

After defining annular apertures, we extracted source spectra from the
target cluster and background spectra from the corresponding
normalized blank-sky dataset. By standard {\textsc{CIAO}} means we
created weighted effective area functions (WARFs) and redistribution
matrices (WRMFs) for each cluster using a flux-weighted map (WMAP)
across the entire extraction region. The WMAP was calculated over the
energy range 0.3-2.0 keV to weight calibrations that vary as a
function of position on the chip. The CCD characteristics which affect
the analysis of extended sources, such as energy dependent vignetting,
are contained within these files. Each spectrum was then binned to
contain a minimum of 25 counts per channel.

\section{Spectral Analysis} \label{sec:specan}

\subsection{Fitting} \label{sec:fitting}

Spectra were fit with {\textsc{XSPEC 11.3.2ag}} \citep{1996ASPC..101...17A}
using a single-temperature {\textsc{MeKaL}} model in combination with the
photoelectric absorption model {\textsc{WABS}} \citep{1983ApJ...270..119M}
to account for Galactic absorption. Galactic absorption values,
$N_{HI}$, are taken from \cite{1990ARA&A..28..215D}. The potentially
free parameters of the absorbed thermal model are
$N_{HI}$, X-ray temperature ($T_{X}$), metal abundance normalized to solar
(elemental ratios taken from \citealt{1989GeCoA..53..197A}), and a
normalization proportional to the integrated emission measure of the
cluster. Results from the fitting are presented in Tables
\ref{tab:r2500specfits} and \ref{tab:r5000specfits}. No
systematic error is added during fitting, and thus all quoted errors
are statistical only. The statistic used during fitting was $\chi^2$
({\textsc{XSPEC}} statistics package \textsc{chi}). Every cluster
analyzed was found to have greater than 1500 background-subtracted
source counts in the spectrum.

For some clusters, more than one observation was available in the
archive. We utilized the power of the combined exposure time by first
extracting independent spectra, WARFs, WRMFs, normalized background
spectra, and soft residuals for each observation. Then, these
independent spectra were read into \textsc{XSPEC} simultaneously and
fit with one spectral model which had all parameters, except
normalization, tied among the spectra. The simultaneous fit is what is
reported for these clusters, denoted by a star ($\star$), in Tables
\ref{tab:r2500specfits} and \ref{tab:r5000specfits}.

Additional statistical error was introduced into the fits because of
uncertainty associated with the soft local background component
discussed in \S\ref{sec:background}. To estimate the sensitivity of
our best-fit temperatures to this uncertainty, we used the differences
between $T_{X}$ for a model using the best-fit soft background
normalization and $T_{X}$ for models using $\pm1\sigma$ of the soft
background normalization. The statistical uncertainty of the original
fit and the additional uncertainty inferred from the range of
normalizations to the soft X-ray background component were then added
in quadrature to produce a final error. In all cases this additional
background error on the temperature was less than 10\% of the total
statistical error, and therefore represents a minor inflation of the
error budget.

When comparing fits with fixed Galactic column density with those
where it was a free parameter, we found that neither the goodness of
fit per free parameter nor the best-fit $T_{X}$ were significantly
different. Thus, $N_{HI}$ was fixed at the Galactic value with the
exception of three cases: Abell 399 \citep{2004MNRAS.351.1439S}, Abell
520, and Hercules A. For these three clusters $N_{HI}$ is a free
parameter. In all fits, the metal abundance was a free parameter.

After fitting we rejected several datasets as their best-fit $T_{2.0-7.0}$
had no upper bound in the 90\% confidence interval and thus were
insufficient for our analysis. All fits for the clusters Abell 781,
Abell 1682, CL J1213+0253, CL J1641+4001, IRAS 09104+4109, Lynx E,
MACS J1824.3+4309, MS 0302.7+1658, and RX J1053+5735 were rejected. We
also removed Abell 2550 from our sample after finding it to be an
anomalously cool ($T_{X} \sim$ 2 keV) ``cluster''. In fact, Abell 2550
is a line-of-sight set of groups, as discussed by
\cite{2004cgpc.sympE..31M}. After these rejections, we are left with
a final sample of 166 clusters which have $R_{2500-\mathrm{CORE}}$
fits and 192 clusters which have $R_{5000-\mathrm{CORE}}$ fits.

\subsection{Simulated Spectra}\label{sec:simulated}

To quantify the effect a second, cooler gas component would have on
the fit of a single-component spectral model, we created an ensemble
of simulated spectra for each real spectrum in our entire sample using
{\textsc{XSPEC}}. With these simulated spectra we sought to answer the
question: Given the count level in each observation of our sample, how
bright must a second temperature component be for it to affect the
observed temperature ratio? Put another way, we asked at what flux
ratio a second gas phase produces a temperature ratio, $T_{HBR}$, of
greater than unity with 90\% confidence.

We began by adding the observation-specific background to a convolved,
absorbed thermal model with two temperature components observed for a
time period equal to the actual observation's exposure time and adding Poisson
noise. For each realization of an observation's simulated spectrum, we
defined the primary component to have the best-fit temperature and
metallicity of the $R_{2500-\mathrm{CORE}}$ 0.7-7.0 keV fit, or
$R_{5000-\mathrm{CORE}}$ if no $R_{2500-\mathrm{CORE}}$ fit was
performed. We then incremented the secondary component temperature
over the values 0.5, 0.75, 1.0, 2.0, and 3.0 keV. The metallicity of
the secondary component was fixed and set equal to the metallicity of
the primary component.

We adjusted the normalization of the simulated two-component spectra to
achieve equivalent count rates to those in the real spectra. The sum
of normalizations can be expressed as $N = N_1 + \xi \cdot N_2$. We
set the secondary component normalization to $N_2 = \xi \cdot N_{bf}$
where $N_{bf}$ is the best-fit normalization of the appropriate
0.7-7.0 keV fit and $\xi$ is a preset factor taking the values 0.4,
0.3, 0.2, 0.15, 0.1, and 0.05. The primary component normalization,
$N_1$, was determined through an iterative process to make real and
simulated spectral count rates match. The parameter $\xi$ therefore
represents the fractional contribution of the cooler component to the
overall count rate.

There are many systematics at work in the full ensemble of observation
specific simulated spectra, such as redshift, column density, and
metal abundance. Thus as a further check of spectral sensitivity to
the presence of a second gas phase, we simulated additional spectra
for the case of an idealized observation. We followed a similar
procedure to that outlined above, but in this instance we used a finer
temperature and $\xi$ grid of $T_2 = 0.5 \rightarrow 3.0$ in steps of
0.25 keV, and $\xi = 0.02 \rightarrow 0.4$ in steps of 0.02. The
input spectral model was $N_{HI} = 3.0\times10^{20}$ cm$^{-2}$, $T_1 = 5$
keV, $Z/Z_{\odot} = 0.3$ and $z = 0.1$. We also varied the exposure
times such that the total number of counts in the 0.7-7.0 keV band was
15K, 30K, 60K, or 120K. For these spectra we used the on-axis sample
response files provided to Cycle 10
proposers\footnote{http://cxc.harvard.edu/caldb/prop\_plan/imaging/index.html}.
Poisson noise is added, but no background is considered.

\begin{figure}
\begin{center}
\includegraphics*[width=\textwidth, trim=0mm 0mm 0mm 0mm, clip]{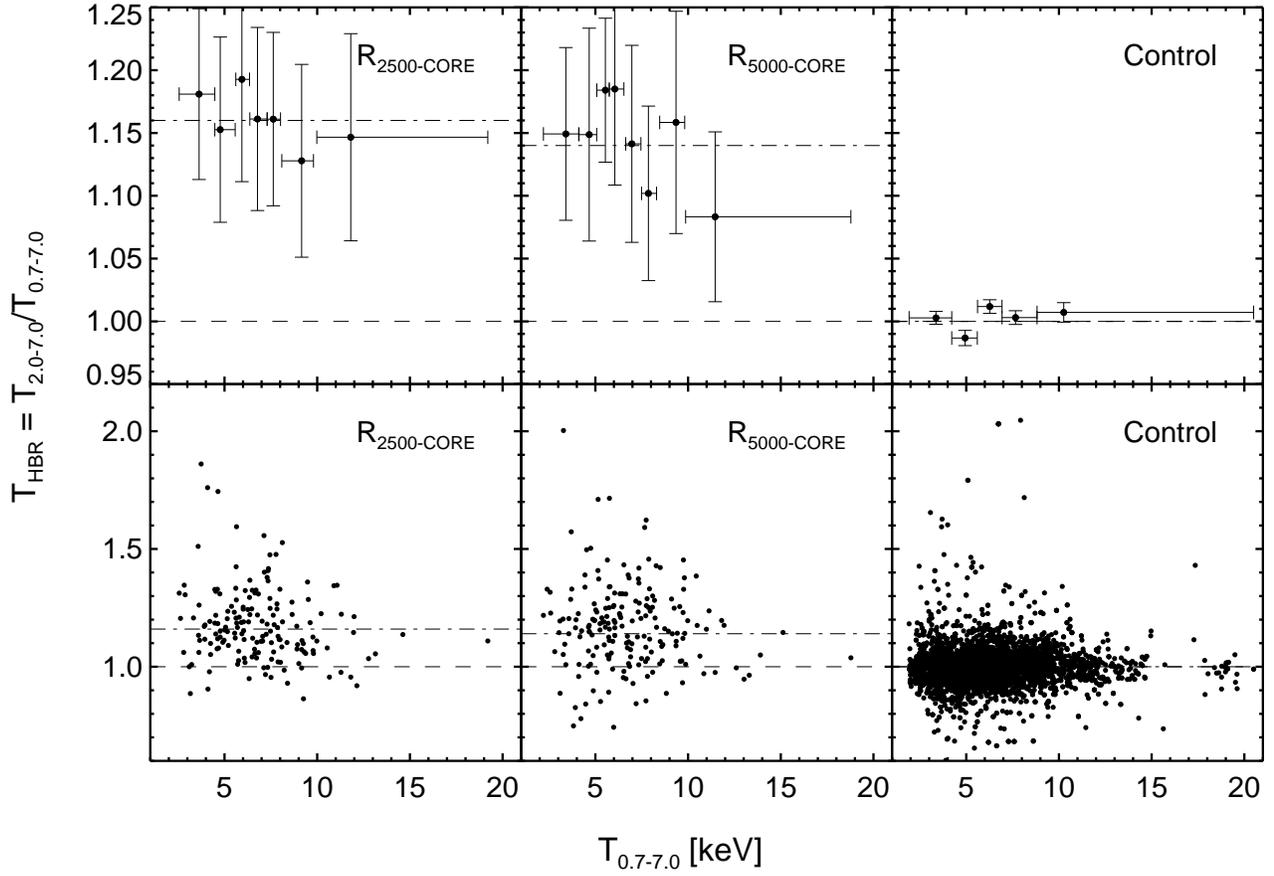}
\caption{
Best-fit temperatures for the hard-band, $T_{2.0-7.0}$, divided by the
broad-band, $T_{0.7-7.0}$, and plotted against the broad-band
temperature. For binned data, each bin contains 25 clusters, with the
exception of the highest temperature bins which contain 16 and 17 for
$R_{2500-\mathrm{CORE}}$ and $R_{5000-\mathrm{CORE}}$, respectively. The
simulated data bins contain 1000 clusters with the last bin having 780
clusters. The line of equality is shown as a dashed line and the
weighted mean for the full sample is shown as a dashed-dotted
line. Error bars are omitted in the unbinned data for clarity. Note
the net skewing of $T_{HBR}$ to greater than unity for both apertures
with no such trend existing in the simulated data. The dispersion of
$T_{HBR}$ for the real data is also much larger than the dispersion of
the simulated data.
}
\label{fig:ftx}
\end{center}
\end{figure}

We also simulated a control sample of single-temperature models. The
control sample is simply a simulated version of the best-fit
model. This control provides us with a statistical test of how often
the actual hard-component temperature might differ from a broad-band
temperature fit if calibration effects are under control. Fits for the
control sample are shown in the far right panels of Figure
\ref{fig:ftx}.

For each observation, we have 65 total simulated spectra: 35
single-temperature control spectra and 30 two-component simulated
spectra (five second temperatures, each with six different $\xi$). Our
resulting ensemble of simulated spectra contains 12,765 spectra. After
generating all the spectra we followed the same fitting routine
detailed in \S\ref{sec:fitting}.

With the ensemble of simulated spectra we then asked the question:
for each $T_2$ and $\Delta T_X$ (defined as the difference between
the primary and secondary temperature components) what is the minimum
value of $\xi$, called $\xi_{min}$, that produces $T_{HBR} \geq 1.1$ at
90\% confidence? From our analysis of these simulated spectra we have
found these important results:
\begin{enumerate}

\item In the control sample, a single-temperature model rarely
($\sim 2\%$ of the time) gives a significantly different $T_{0.7-7.0}$ and
$T_{2.0-7.0}$. The weighted average (right panels of
Fig. \ref{fig:ftx}) for the control sample is $1.002 \pm 0.001$ and the
standard deviation is $\pm0.044$. The $T_{HBR}$ distribution for the
control sample appears to have an intrinsic width which is likely
associated with statistical noise of fitting in {\textsc{XSPEC}}
(Dupke, private communication). This result indicates that our
remaining set of observations is statistically sound, e.g. our finding
that $T_{HBR}$ significantly differs from 1.0 cannot result from
statistical fluctuations alone.

\item Shown in Table \ref{tab:simres} are the contributions
a second cooler component must make in the case of the idealized
spectra in order to produce $T_{HBR} \geq 1.1$ at 90\% confidence. In
general, the contribution of cooler gas must be $> 10\%$ for $T_2 < 2$
keV to produce $T_{HBR}$ as large as 1.1. The increase in percentages
at $T_2 < 1.0$ keV is owing to the energy band we consider (0.7-7.0
keV) as gas cooler than 0.7 keV must be brighter than at 1.0 keV in
order to make an equivalent contribution to the soft end of the
spectrum at 0.7 keV.

\item In the full ensemble of observation-specific simulated spectra,
we find a great deal of statistical scatter in $\xi_{min}$ at any given
$\Delta T_X$. This was expected as the full ensemble is a
superposition of spectra with a broad range of total counts, $N_{HI}$,
redshifts, abundance, and backgrounds. But using the idealized
simulated spectra as a guide, we find for those spectra with
$N_{\mathrm{counts}} \gtrsim 15000$, producing $T_{HBR} \geq 1.1$ at
90\% confidence again requires the cooler gas to be contributing $>
10\%$ of the emission. These results are also summarized in Table
\ref{tab:simres}. The good agreement between the idealized and
observation-specific simulated spectra indicates that while many more
factors are in play for the observation-specific spectra, they do not
degrade our ability to reliably measure $T_{HBR} > 1.1$.
The trend here of a common soft component sufficient to change the
temperature measurement in a single-temperature model is statistical,
a result that comes from an aggregate view of the sample rather than
any individual fit.

\item As redshift increases, gas cooler than 1.0 keV is slowly
redshifted out of the observable X-ray band. As expected, we find from
our simulated spectra that for $z \geq 0.6$, $T_{HBR}$ is no longer
statistically distinguishable from unity. In addition, the
$T_{2.0-7.0}$ lower boundary nears convergence with the $T_{0.7-7.0}$
lower boundary as $z$ increases, and for $z = 0.6$, the hard-band
lower limit is 1.25 keV, while at the highest redshift considered, $z =
1.2$, the hard-band lower limit is only 0.91 keV. For the 14
clusters with $z \geq 0.6$ in our real sample we are most likely
underestimating the actual amount of temperature inhomogeneity. We
have tested the effect of excluding these clusters on our results, and
find a negligible change in the overall skew of $T_{HBR}$ to greater
than unity.
\end{enumerate}

\section{Results and Discussion} \label{sec:r&d}

\subsection{Temperature Ratios} \label{sec:tfresults}

For each cluster we have measured a ratio of the hard-band
to broad-band temperature defined as $T_{HBR}$ =
$T_{2.0-7.0}$/$T_{0.7-7.0}$. We find that the mean $T_{HBR}$ for our
entire sample is greater than unity at more than $12\sigma$
significance. The weighted mean values for our sample are shown in
Table \ref{tab:wavg}. Presented in Figure  \ref{fig:ftx} are the binned
weighted means and raw $T_{HBR}$ values for $R_{2500-\mathrm{CORE}}$,
$R_{5000-\mathrm{CORE}}$, and the simulated control sample. The peculiar
points with $T_{HBR} <$ 1 are all statistically consistent with
unity. The presence of clusters with $T_{HBR}$ = 1 suggests that
systematic calibration uncertainties are not the sole reason for
deviations of $T_{HBR}$ from 1. We also find that the temperature
ratio does not depend on the best-fit broad-band temperature, and that
the observed dispersion of $T_{HBR}$ is greater than the predicted
dispersion arising from systematic uncertainties.

The uncertainty associated with each value of $T_{HBR}$ is dominated by
the larger error in $T_{2.0-7.0}$, and on average, $\Delta T_{2.0-7.0} \approx
2.3\Delta T_{0.7-7.0}$. This error interval discrepancy naturally results
from excluding the bulk of a cluster's emission which occurs below 2
keV. While choosing a temperature-sensitive cut-off energy for the
hard-band (other than 2.0 keV) might maintain a more consistent
error budget across our sample, we do not find any systematic trend in
$T_{HBR}$ or the associated errors with cluster temperature.

\subsection{Systematics} \label{sec:sys}

In this study we have found the average value of $T_{HBR}$ is significantly
greater than one and that $\sigma_{HBR} > \sigma_{\mathrm{control}}$, with the
latter result being robust against systematic uncertainties. As
predicted by ME01, both of these results are expected to arise
naturally from the hierarchical formation of clusters. But systematic
uncertainty related to {\it Chandra} instrumentation or other sources could
shift the average value of $T_{HBR}$ one would get from ``perfect'' data. In
this section we consider some additional sources of uncertainty.

First, the disagreement between {\it XMM-Newton} and {\it Chandra}
cluster temperatures has been noted in several independent studies,
i.e. \cite{2005ApJ...628..655V} and \cite{chanxmmdis}. But the source of this
discrepancy is not well understood and efforts to perform
cross-calibration between {\it XMM-Newton} and {\it Chandra} have thus
far not been conclusive. One possible explanation is poor calibration
of {\it Chandra} at soft X-ray energies which may arise from a
hydrocarbon contaminant on the High Resolution Mirror Assembly (HRMA)
similar in nature to the contaminant on the ACIS detectors
\citep{aciscontaminant}. We have assessed this possibility by looking
for systematic trends in $T_{HBR}$ with time or temperature, as such a
contaminant would most likely have a temperature and/or time
dependence.

\begin{figure}
\begin{center}
\includegraphics*[width=\textwidth, trim=0mm 0mm 0mm 0mm, clip]{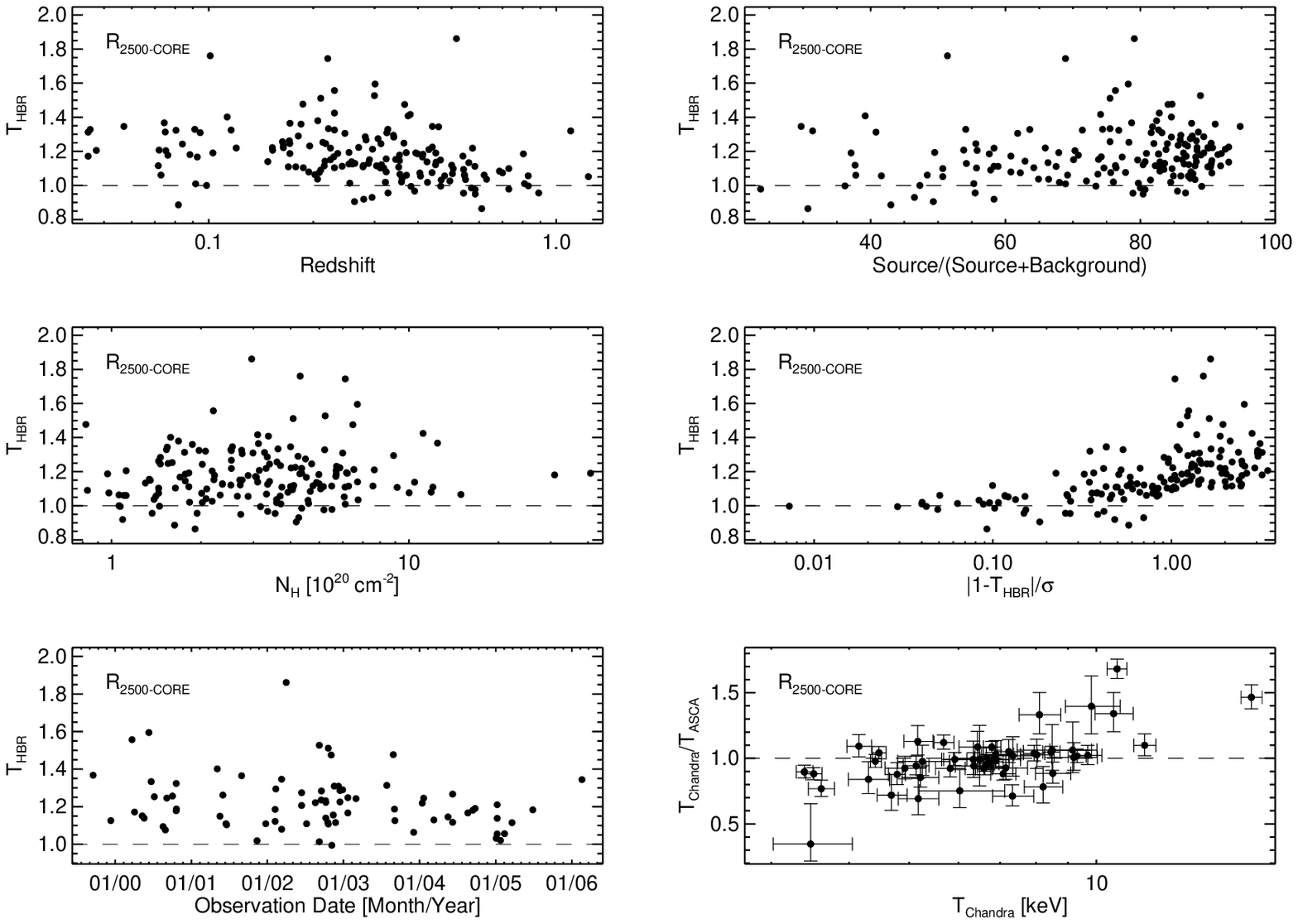}
\caption{
Plotted here are a few possible sources of systematic uncertainty
versus $T_{HBR}$ calculated for the $R_{2500-\mathrm{CORE}}$
apertures (166 clusters). Error bars have been omitted in several
plots for clarity. The line of equality is shown as a dashed line in
all panels.
{\bfseries\em{(Upper-left:)}} $T_{HBR}$ versus redshift for the
entire sample. The trend in $T_{HBR}$ with redshift is expected as the
$T_{2.0-7.0}$ lower boundary nears convergence with the $T_{0.7-7.0}$
lower boundary at $z \approx 1.85$. Weighted values of $T_{HBR}$ are
consistent with unity starting at $z \sim
0.6$.
{\bfseries\em{(Upper-right:)}} $T_{HBR}$ versus percentage of
spectrum flux which is attributed to the source. We find no trend with
signal-to-noise which suggests calibration uncertainty not is playing
a major role in our results.
{\bfseries\em{(Middle-left:)}} $T_{HBR}$
versus Galactic column density. We find no trend in absorption which
would result if $N_{HI}$ values are inaccurate or if we had improperly
accounted for local soft
contamination.
{\bfseries\em{(Middle-right:)}} $T_{HBR}$ versus the
deviation from unity in units of measurement uncertainty. Recall that
we have used 90\% confidence ($1.6\sigma$) for our analysis.
{\bfseries\em{(Lower-left:)}} $T_{HBR}$ plotted versus
observation start date. The plotted points are culled from the full
sample and represent only clusters which have a single observation and
where the spectral flux is $> 75\%$ from the source. We note no
systematic trend with time.
{\bfseries\em{(Lower-right:)}} Ratio of {\it Chandra}
temperatures derived in this work to {\it ASCA} temperatures taken
from Don Horner's thesis. We note a trend of comparatively hotter {\it
Chandra} temperatures for clusters $> 10$ keV, otherwise our derived
temperatures are in good agreement with those of {\it ASCA}.
}
\label{fig:sysr25}
\end{center}
\end{figure}

\begin{figure}
\begin{center}
\includegraphics*[width=\textwidth, trim=0mm 0mm 0mm 0mm, clip]{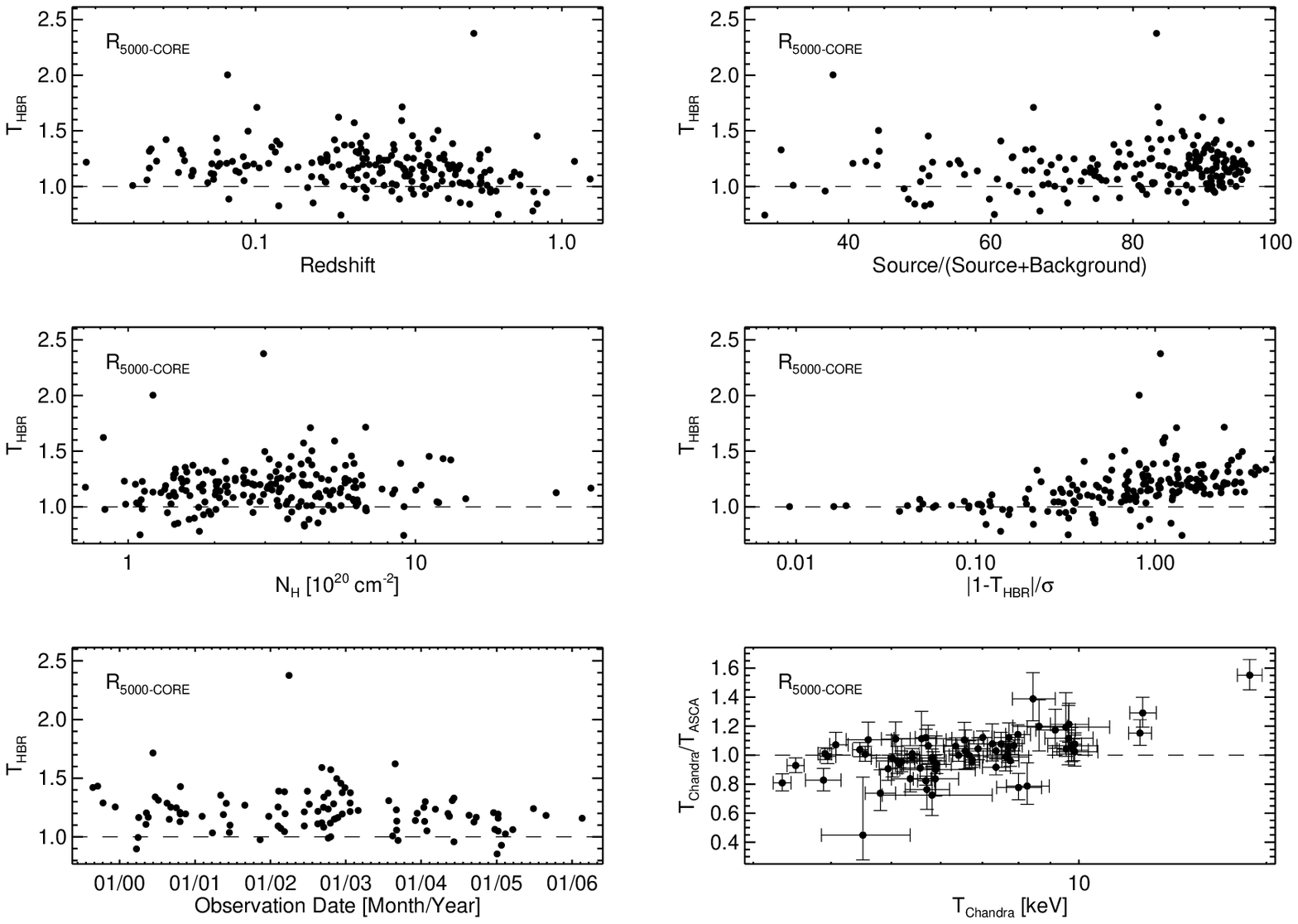}
\caption{
Plotted here are a few possible sources of systematic uncertainty
versus $T_{HBR}$ calculated for the $R_{5000-\mathrm{CORE}}$
apertures (192 clusters). Error bars have been omitted in several plots for
clarity. The line of equality is shown as a dashed line in all
panels.
{\bfseries\em{(Upper-left:)}} $T_{HBR}$ versus redshift for the
entire sample. The trend in $T_{HBR}$ with redshift is expected as the
$T_{2.0-7.0}$ lower boundary nears convergence with the $T_{0.7-7.0}$
lower boundary at $z \approx 1.85$. Weighted values of $T_{HBR}$ are
consistent with unity starting at $z \sim
0.6$.
{\bfseries\em{(Upper-right:)}} $T_{HBR}$ versus percentage of
spectrum flux which is attributed to the source. We find no trend with
signal-to-noise which suggests calibration uncertainty is not playing
a major role in our results.
{\bfseries\em{(Middle-left:)}} $T_{HBR}$
versus Galactic column density. We find no trend in absorption which
would result if $N_{HI}$ values are inaccurate or if we had improperly
accounted for local soft
contamination.
{\bfseries\em{(Middle-right:)}} $T_{HBR}$ versus the
deviation from unity in units of measurement uncertainty. Recall that
we have used 90\% confidence ($1.6\sigma$) for our analysis.
{\bfseries\em{(Lower-left:)}} $T_{HBR}$ plotted versus
observation start date. The plotted points are culled from the full
sample and represent only clusters which have a single observation and
where the spectral flux is $> 75\%$ from the source. We note no
systematic trend with time.
{\bfseries\em{(Lower-right:)}} Ratio of {\it Chandra}
temperatures derived in this work to {\it ASCA} temperatures taken
from Don Horner's thesis. We note a trend of comparatively hotter {\it
Chandra} temperatures for clusters $> 10$ keV, otherwise our derived
temperatures are in good agreement with those of {\it ASCA}.
}
\label{fig:sysr50}
\end{center}
\end{figure}

As noted in \S\ref{sec:tfresults} and seen in Figure \ref{fig:ftx}, we
find no systematic trend with temperature either for the full sample
or for a sub-sample of single-observation clusters with $> 75\%$ of
the observed flux attributable to the source (higher signal-to-noise
observations will be more affected by calibration
uncertainty). Plotted in the lower-left pane of Figures
\ref{fig:sysr25} and \ref{fig:sysr50} is $T_{HBR}$ versus time for
single observation clusters (clusters with multiple observations are
fit simultaneously and any time effect would be washed out) where
the spectral flux is $> 75\%$ from the source. We find no significant
systematic trend in $T_{HBR}$ with time, which suggests that if
$T_{HBR}$ is affected by any contamination of {\it Chandra}'s HRMA,
then the contaminant is most likely not changing with time. Our
conclusion on this matter is that the soft calibration uncertainty is
not playing a dominant role in our results.

Aside from instrumental and calibration effects, some other possible
sources of systematic error are signal-to-noise (S/N), redshift
selection, Galactic absorption, and metallicity. Also presented in
Figures \ref{fig:sysr25} and \ref{fig:sysr50} are three of these
parameters versus $T_{HBR}$ for $R_{2500-\mathrm{CORE}}$ and
$R_{5000-\mathrm{CORE}}$, respectively. The trend in $T_{HBR}$ with
redshift is expected as the 2.0/(1+$z$) keV hard-band lower boundary
nears convergence with the 0.7 keV broad-band lower boundary at $z
\approx 1.85$. We find no systematic trends of $T_{HBR}$ with S/N or
Galactic absorption, which might occur if the skew in $T_{HBR}$ were a
consequence of poor count statistics, inaccurate Galactic absorption,
or very poor calibration. In addition, the ratio of $T_{HBR}$ for
$R_{2500-\mathrm{CORE}}$ to $R_{5000-\mathrm{CORE}}$ for every cluster
in our sample does not significanlty deviate from unity. Our results
are robust to changes in aperture size.

Also shown in Figures \ref{fig:sysr25} and \ref{fig:sysr50} are the
ratios of {\it ASCA} temperatures taken from Don Horner's thesis to
{\it Chandra} temperatures derived in this work. The spurious point
below 0.5 with very large error bars is MS 2053.7-0449, which has a
poorly constrained {\it ASCA} temperature of
$10.03^{+8.73}_{-3.52}$. Our value of $\sim 3.5$ keV for this cluster
is in agreement with the recent work of
\cite{2007astro.ph..3156M}. Not all our sample clusters have an {\it
ASCA} temperature, but a sufficient number (53) are available to make
this comparison reliable. Apertures used in the extraction of {\it
ASCA} spectra had no core region removed and were substantially larger
than $R_{2500}$. {\it ASCA} spectra were also fit over a broader
energy range (0.6-10 keV) than we use here. Nonetheless, our
temperatures are in good agreement with those from {\it ASCA}, but we
do note a trend of comparatively hotter {\it Chandra} temperatures for
$T_{Chandra} > 10$ keV. For both apertures, the clusters with
$T_{Chandra} > 10$ keV are Abell 1758, Abell 2163, Abell 2255, and RX
J1347.5-1145. Based on this trend, we test excluding the hottest
clusters ($T_{Chandra} > 10$ keV where {\it ASCA} and {\it Chandra}
disagree) from our sample. The mean temperature ratio for
$R_{2500-\mathrm{CORE}}$ remains $1.16$ and the error of the mean
increases from $\pm 0.014$ to $\pm 0.015$, while for
$R_{5000-\mathrm{CORE}}$ $T_{HBR}$ increases by a negligible $0.9\%$
to $1.15\pm 0.014$. Our results are not being influenced by the
inclusion of hot clusters.

\begin{figure}
\begin{center}
\includegraphics*[width=\textwidth, trim=0mm 0mm 0mm 0mm, clip]{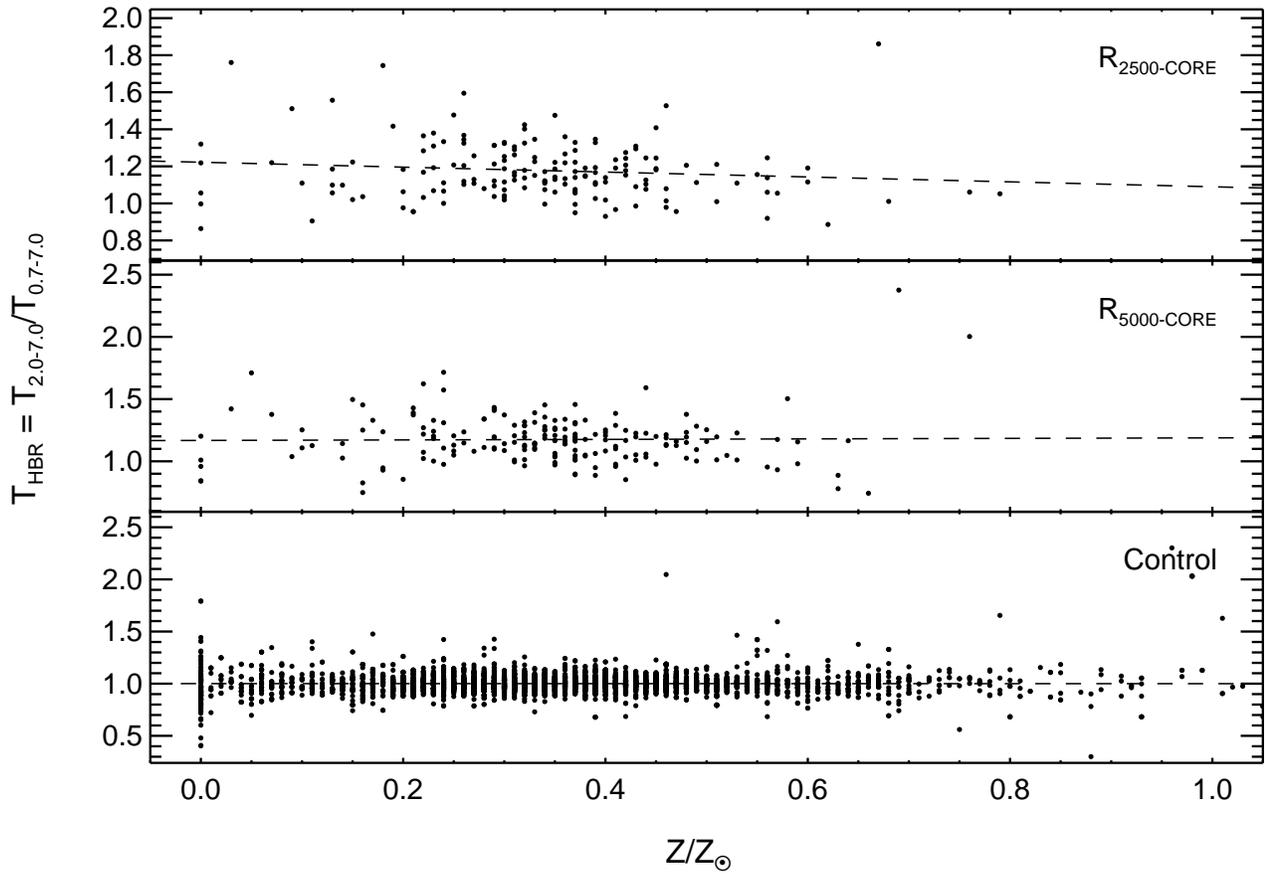}
\caption{
Plotted here is $T_{HBR}$ as a function of metal abundance for 
$R_{2500-\mathrm{CORE}}$, $R_{5000-\mathrm{CORE}}$, and the Control
sample (see discussion of control sample in
\S\ref{sec:simulated}). Error bars are omitted for clarity. The
dashed-line represents the linear best-fit using the bivariate
correlated error and intrinsic scatter (BCES) method of
\cite{1996ApJ...470..706A} which takes into consideration errors on
both $T_{HBR}$ and abundance when performing the fit. We note no trend
in $T_{HBR}$ with metallicity (the apparent trend in the top panel is
not significant) and also note the low dispersion in the control
sample relative to the observations. The striation of abundance arises
from our use of two decimal places in recording the best-fit values
from {\textsc{XSPEC}}.
}
\label{fig:metal}
\end{center}
\end{figure}

The temperature range of the clusters we've analyzed ($T_X \sim 3-20$
keV) is broad enough that the effect of metal abundance on the
inferred spectral temperature is clearly not negligible. In Figure
\ref{fig:metal} we have plotted $T_{HBR}$ versus abundance in solar
units. Despite covering a factor of seven in temperature and
metal abundances ranging from $Z/Z_{\odot} \approx 0$ to solar, we
find no trend in $T_{HBR}$ with metallicity. The slight trend in the
$R_{2500-\mathrm{CORE}}$ aperture (top panel of Figure
\ref{fig:metal}) is insignificant, while there is no trend at all in
the control sample or $R_{5000-\mathrm{CORE}}$ aperture.

\subsection{Using $T_{HBR}$ as a Test of Relaxation} \label{sec:relax}

\subsubsection{Cool Core Versus Non-Cool Core}\label{sec:ccncc}

As discussed in \S\ref{sec:intro}, ME01 gives us reason to believe the
observed skewing of $T_{HBR}$ to greater than unity is related to the
dynamical state of a cluster. It has also been suggested that the
process of cluster formation and relaxation may robustly result in
the formation of a cool core
\citep{2006ApJ...640..673O,2007arXiv0708.1954B}. Depending upon
classification criteria, completeness, and possible selection biases,
studies of flux-limited surveys have placed the prevalence of cool
cores at $34-60\%$ \citep{1997MNRAS.292..419W, 1998MNRAS.298..416P,
2005MNRAS.359.1481B, 2007A&A...466..805C}. It has thus become rather
common to divide up the cluster population into two distinct classes,
cool core (CC) and non-cool core (NCC), for the purpose of discussing
their different formation or merger histories. We thus sought to
identify which clusters in our sample have cool cores, which do not,
and if the presence or absence of a cool core is correlated with
$T_{HBR}$. It is very important to recall that we excluded the core
during spectral extraction and analysis.

To classify the core of each cluster, we extracted a spectrum for the
50 kpc region surrounding the cluster center and then defined a
temperature decrement,
\begin{equation}
T_{\mathrm{dec}} = T_{50}/T_{\mathrm{cluster}}
\label{eqn:tdec}
\end{equation}
where $T_{50}$ is the temperature of the inner 50 kpc and
$T_{\mathrm{cluster}}$ is either the $R_{2500-\mathrm{CORE}}$ or
$R_{5000-\mathrm{CORE}}$ temperature. If $T_{\mathrm{dec}}$ was
2$\sigma$ less than unity, we defined the cluster as having a CC,
otherwise the cluster was defined as NCC. We find CCs in 35\% of our
sample and when we lessen the significance needed for CC
classification from 2$\sigma$ to 1$\sigma$, we find 46\% of our sample
clusters have CCs. It is important to note that the frequency of CCs
in our study is consistent with other more detailed studies of CC/NCC
populations.

When fitting for $T_{50}$, we altered the method outlined
in \S\ref{sec:fitting} to use {\textsc{XSPEC}}'s modified Cash statistic
\citep{1979ApJ...228..939C}, {\textsc{cstat}}, on ungrouped
spectra. This choice was made because the distribution of counts per
bin in low count spectra is not Gaussian but instead
Poisson. As a result, the best-fit temperature using $\chi^2$ is
typically cooler \citep{1989ApJ...342.1207N, 2007A&A...462..429B}. We
have explored this systematic in {\bfseries\em{all}} of our fits and
found it to be significant only in the lowest count spectra of the
inner 50 kpc apertures discussed here. But, for consistency, we fit all
inner 50 kpc spectra using the modified Cash statistic.

\begin{figure}
\begin{center}
\includegraphics*[width=\textwidth, trim=15mm 10mm 0mm 0mm, clip]{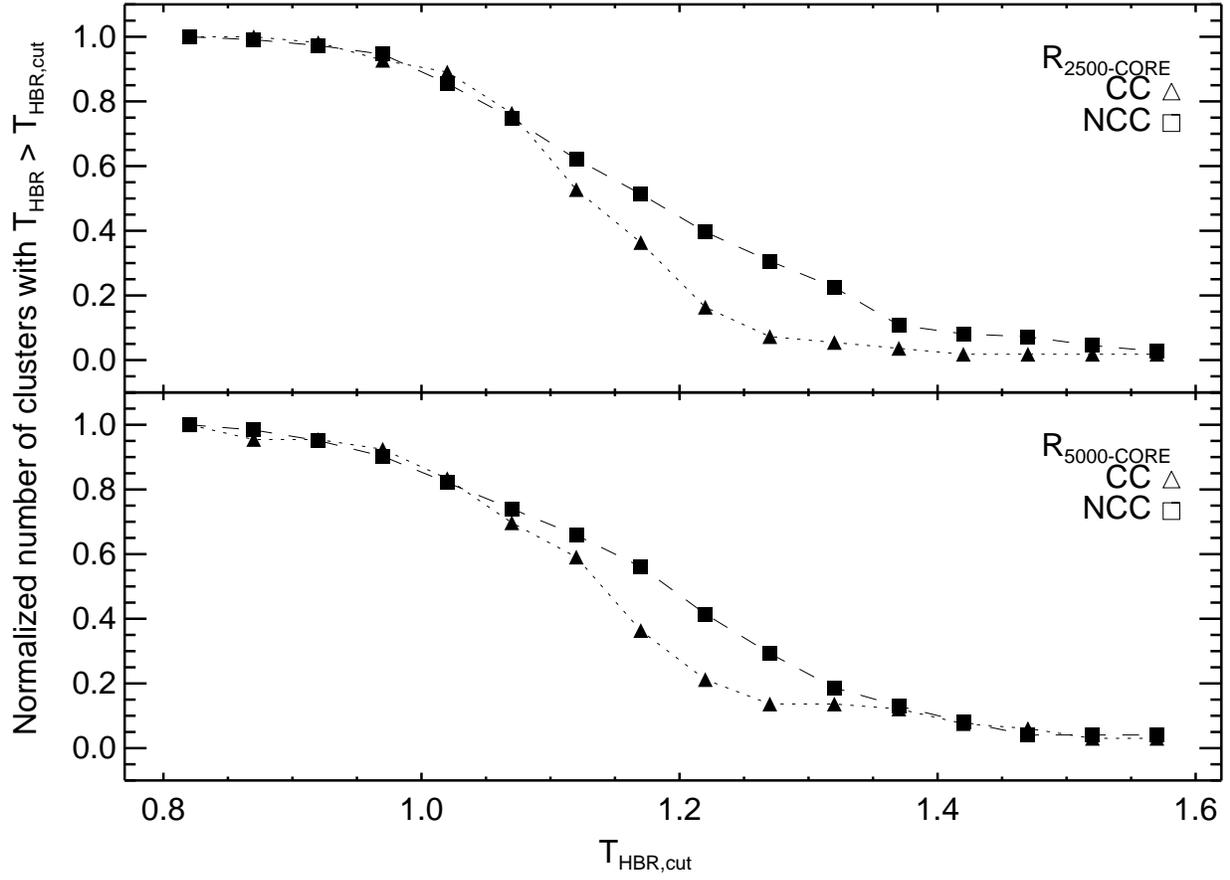}
\caption{
Plotted here is the normalized number of cool core (CC) and non-cool core
(NCC) clusters as a function of cuts in $T_{HBR}$. There are 166
clusters plotted in the top panel and 192 in the bottom panel. We have
defined a cluster as having a cool core (CC) when the temperature for
the 50 kpc region around the cluster center divided by the temperature
for $R_{2500-\mathrm{CORE}}$, or $R_{5000-\mathrm{CORE}}$, was less
than one at the $2\sigma$ level. We then take cuts in $T_{HBR}$ at the
$1\sigma$ level and ask how many CC and NCC clusters are above these
cuts. The number of CC clusters falls off more rapidly than NCC
clusters in this classification scheme suggesting higher values of
$T_{HBR}$ prefer less relaxed systems which do not have cool
cores. This result is insensitive to our choice of significance level
in both the core classification and $T_{HBR}$ cuts.
}
\label{fig:cc_ncc_bin}
\end{center}
\end{figure}

With each cluster core classified, we then took cuts in $T_{HBR}$ 
and asked how many CC and NCC clusters were above these cuts. 
Figure \ref{fig:cc_ncc_bin} shows the normalized number of CC and NCC
clusters as a function of cuts in $T_{HBR}$. If $T_{HBR}$ were
insensitive to the state of the cluster core, we expect, for normally
distributed $T_{HBR}$ values, to see the number of CC and NCC clusters
decreasing in the same way. However, the number of CC clusters falls
off more rapidly than the number of NCC clusters. If the presence of a
CC is indicative of a cluster's advancement towards complete
virialization, then the significantly steeper decline in the percent
of CC clusters versus NCC as a function of increasing $T_{HBR}$
indicates higher values of $T_{HBR}$ are associated with a less
relaxed state. This result is insensitive to our choice of
significance level in the core classification, i.e. the result is the
same whether using $1\sigma$ or $2\sigma$ significance when
considering $T_{\mathrm{dec}}$.

Because of the CC/NCC definition we selected, our identification of
CCs and NCCs was only as robust as the errors on $T_{50}$ allowed. One can
thus ask the question, did our definition bias us towards
finding more NCCs than CCs? To explore this question we simulated 20
spectra for each observation following the method outlined in
\S\ref{sec:simulated} for the control sample but using the inner 50
kpc spectral best-fit values as input. For each simulated spectrum, we
calculated a temperature decrement (Eqn. \ref{eqn:tdec}) and
re-classified the cluster as having a CC or NCC. Using the new set of
mock classifications we assigned a reliability factor, $\psi$, to each
real classification, which is simply the fraction of mock
classifications which agree with the real classification. A value of
$\psi = 1.0$ indicates complete agreement, and $\psi = 0.0$ indicating no
agreement. When we removed clusters with $\psi < 0.9$ and repeated the
analysis above, we found no significant change in the trend of a
steeper decrease in the relative number of CC versus NCC clusters as a
function of $T_{HBR}$.

Recall that the coolest ICM gas is being redshifted out of the
observable band as $z$ increases and becomes a significant effect at
$z \geq 0.6$ (\S\ref{sec:simulated}). Thus, we are likely not detecting
``weak'' CCs in the highest redshift clusters of our sample and
consequently these cores are classified as NCCs and are artificially
increasing the NCC population. When we excluded the 14 clusters at $z
\geq 0.6$ from this portion of our analysis and repeated the
calculations, we found no significant change in the results.

\subsubsection{Mergers Versus Non-Mergers}\label{sec:merge}

Looking for a correlation between cluster relaxation and a skewing in
$T_{HBR}$ was the primary catalyst of this work. The result that
increasing values of $T_{HBR}$ are more likely to be associated with
clusters harboring non-cool cores gives weight to that hypothesis. But,
the simplest relation to investigate is if $T_{HBR}$ is preferentially
higher in merger systems. Thus, we now discuss clusters with the
highest significant values of $T_{HBR}$ and attempt to establish, via
literature based results, the dynamic state of these systems.

The subsample of clusters on which we focus have a $T_{HBR} > 1.1$ at
90\% confidence for both their $R_{2500-\mathrm{CORE}}$ and
$R_{5000-\mathrm{CORE}}$ apertures. These clusters are listed in Table
\ref{tab:tf11} and are sorted by the lower limit of $T_{HBR}$. The
clusters with only a $R_{5000-\mathrm{CORE}}$ analysis are listed
separately at the bottom of the table. All 33 clusters listed have a
core classification of $\psi > 0.9$ (see \S\ref{sec:ccncc}). The
choice of the $T_{HBR} > 1.1$ threshold was arbitrary and intended to
limit the number of clusters to which we pay individual attention, but
which is still representative of mid- to high-$T_{HBR}$ values. Only
two clusters -- Abell 697 and MACS J2049.9-3217 -- do not have a
$T_{HBR} > 1.1$ in one aperture and not the other. In both cases
though, this was the result of the lower boundary narrowly missing the
cut, but both clusters still have $T_{HBR}$ significantly greater than
unity.

For those clusters which have been individually studied, they are
listed as mergers based on the conclusions of the literature
authors (cited in Table \ref{tab:tf11}). Many different techniques
were used to determine if a system is a merger:  bimodal galaxy
velocity distributions, morphologies, highly asymmetric temperature
distributions, ICM substructure correlated with subclusters, or
disagreement of X-ray and lensing masses. From Table \ref{tab:tf11} we
can see clusters exhibiting the highest significant values of
$T_{HBR}$ tend to be ongoing or recent mergers. At the 2$\sigma$
level, we find increasing values of $T_{HBR}$ favor merger systems
with NCCs over relaxed, CC clusters. It appears mergers have left a
spectroscopic imprint on the ICM which was predicted by ME01 and which
we observe in our sample.

Of the 33 clusters with $T_{HBR}$ significantly $> 1.1$, only seven
have CCs. Three of those -- MKW3S, 3C 28.0, and RX J1720.1+2638 -- have
their apertures centered on the bright, dense cores in confirmed
mergers. Two more clusters -- Abell 2384 and RX J1525+0958 -- while not
confirmed mergers, have morphologies which are consistent with
powerful ongoing mergers. Abell 2384 has a long gas tail extending
toward a gaseous clump which we assume has recently passed through the
cluster. RXJ1525 has a core shaped like a rounded arrowhead and is
reminiscent of the bow shock seen in 1E0657-56. Abell 907 has no signs
of being a merger system, but the highly compressed surface brightness
contours to the west of the core are indicative of a prominent cold
front, a tell-tale signature of a subcluster merger event
\citep{2007PhR...443....1M}. Abell 2029 presents a very interesting
and curious case because of its seemingly high state of relaxation and
prominent cool core. There are no complementary indications it has
experienced a merger event. Yet its core hosts a wide-angle tail
radio source. It has been suggested that such sources might be
attributable to cluster merger activity
\citep{2000MNRAS.311..649S}. Moreover, the X-ray isophotes to the west
of the bright, peaked core are slightly more compressed and may be an
indication of past gas sloshing resulting from the merger of a small
subcluster. Both of these features have been noted previously,
specifically by \cite{2004ApJ...616..178C, 2005xrrc.procE7.08C}. We
suggest the elevated $T_{HBR}$ value for this cluster lends more
weight to the argument that A2029 has indeed experienced a merger
recently, but how long ago we do not know.

The remaining systems we could not verify as mergers -- RX
J0439.0+0715, MACS J2243.3-0935, MACS J0547.0-3904, Zwicky 1215, MACS
J2311+0338, Abell 267, and NGC 6338 -- have NCCs and X-ray
morphologies consistent with an ongoing or post-merger scenario. Abell
1204 shows no signs of recent or ongoing merger activity; however, it
resides at the bottom of the arbitrary $T_{HBR}$ cut, and as evidenced
by Abell 401 and Abell 1689, exceptional spherical symmetry is no
guarantee of relaxation. Our analysis here is partially at the mercy
of morphological assessment, and only a more stringent study of a
carefully selected subsample or analysis of simulated clusters can
better determine how closely correlated $T_{HBR}$ is with the timeline
of merger events.

\section{Summary and Conclusions}\label{sec:summary}

We have explored the band dependence of the inferred X-ray temperature
of the ICM for 192 well-observed ($N_{counts} > 1500$) clusters of
galaxies selected from the {\it Chandra} Data Archive.

We extracted spectra from the annulus between $R=70$ kpc and
$R=R_{2500}$, $R_{5000}$ for each cluster. We compared the X-ray
temperatures inferred for single-component fits to global spectra
when the energy range of the fit was 0.7-7.0 keV (broad) and when the
energy range was $2.0/(1+z)$-7.0 keV (hard). We found that, on
average, the hard-band temperature is significantly higher than
the broad-band temperature. For the $R_{2500-\mathrm{CORE}}$ aperture
we measured a weighted average of $T_{HBR} = 1.16$ with $\sigma = \pm
0.10$ and $\sigma_{mean} = \pm 0.01$ for the $R_{5000-\mathrm{CORE}}$
aperture, and $T_{HBR} = 1.14$ with $\sigma = \pm 0.12$ and
$\sigma_{mean} = \pm 0.01$. We also found no systematic trends in the
value of $T_{HBR}$, or the dispersion of $T_{HBR}$, with
signal-to-noise, redshift, Galactic absorption, metallicity,
observation date, or broad-band temperature.

In addition, we simulated an ensemble of 12,765 spectra which
contained observation-specific and idealized two-temperature component
models, plus a control sample of single-temperature models. From
analysis of these simulations we found the statistical fluctuations
for a single temperature model are inadequate to explain the 
significantly different $T_{0.7-7.0}$ and $T_{2.0-7.0}$ we measure in
our sample. We also found that the observed scatter, $\sigma_{HBR}$,
is consistent with the presence of unresolved cool ($T_X < 2.0$ keV)
gas contributing a minimum of $>10\%$ of the total emission. The
simulations also show the measured observational scatter in $T_{HBR}$
is greater than the statistical scatter, $\sigma_{control}$. These
results are consistent with the process of hierarchical cluster
formation.

Upon further exploration, we found that $T_{HBR}$ is enhanced
preferentially for clusters which are known merger systems and for
clusters without cool cores. Clusters with temperature decrements in
their cores (known as cool-core clusters) tend to have best-fit
hard-band temperatures that are consistently closer to their
best-fit broad-band temperatures. The correlation of $T_{HBR}$ with
the type of cluster core is insensitive to our choice of
classification scheme and is robust against redshift effects. Our
results qualitatively support the finding by ME01 that the temperature
ratio, $T_{HBR}$, might therefore be useful for statistically
quantifying the degree of cluster relaxation/virialization.

An additional robust test of the ME01 finding should be made with
simulations by tracking $T_{HBR}$ during hierarchical assembly of a
cluster. If $T_{HBR}$ is tightly correlated with a cluster's degree of
relaxation, then it, along with other methods of substructure measure,
may provide a powerful metric for predicting (and therefore reducing)
a cluster's deviation from mean mass-scaling relations. The
task of reducing scatter in scaling relations will be very important
if we are to reliably and accurately measure the mass of clusters.

\acknowledgements
Kenneth Cavagnolo was supported in this work by the National
Aeronautics and Space Administration through {\it Chandra} X-ray
Observatory Archive grants AR-6016X and AR-4017A, with additional
support from a start-up grant for Megan Donahue from Michigan State
University. Megan Donahue and Michigan State University acknowledge
support from the NASA LTSA program NNG-05GD82G. Mark Voit thanks NASA
for support through theory grant NNG-04GI89G. The {\it Chandra}
X-ray Observatory Center is operated by the Smithsonian Astrophysical
Observatory for and on behalf of the National Aeronautics Space
Administration under contract NAS8-03060. This research has made use
of software provided by the {\it Chandra} X-ray Center (CXC) in the
application packages {\textsc{CIAO}}, {\textsc{ChIPS}}, and
{\textsc{Sherpa}}. We thank Alexey Vikhlinin for helpful insight and
expert advice. KWC also thanks attendees of the ``Eight Years of Science
with {\it Chandra} Calibration Workshop'' for stimulating discussion
regarding {\it XMM}-{\it Chandra} cross-calibration.  KWC especially
thanks Keith Arnaud for personally providing support and advice for
mastering {\textsc{XSPEC}}. This research has made use of the
NASA/IPAC Extragalactic Database (NED) which is operated by the Jet
Propulsion Laboratory, California Institute of Technology, under
contract with the National Aeronautics and Space Administration. This
research has also made use of NASA's Astrophysics Data System. {\it
ROSAT} data and software were obtained from the High Energy
Astrophysics Science Archive Research Center (HEASARC), provided by
NASA's Goddard Space Flight Center.

\clearpage
\clearpage

\clearpage
\clearpage

\clearpage
\clearpage

\clearpage
\clearpage

\clearpage
\clearpage

\clearpage
\clearpage

\clearpage
\clearpage

\clearpage
\begin{figure}
\begin{center}
\includegraphics*[width=\textwidth, trim=15mm 10mm 0mm 0mm, clip]{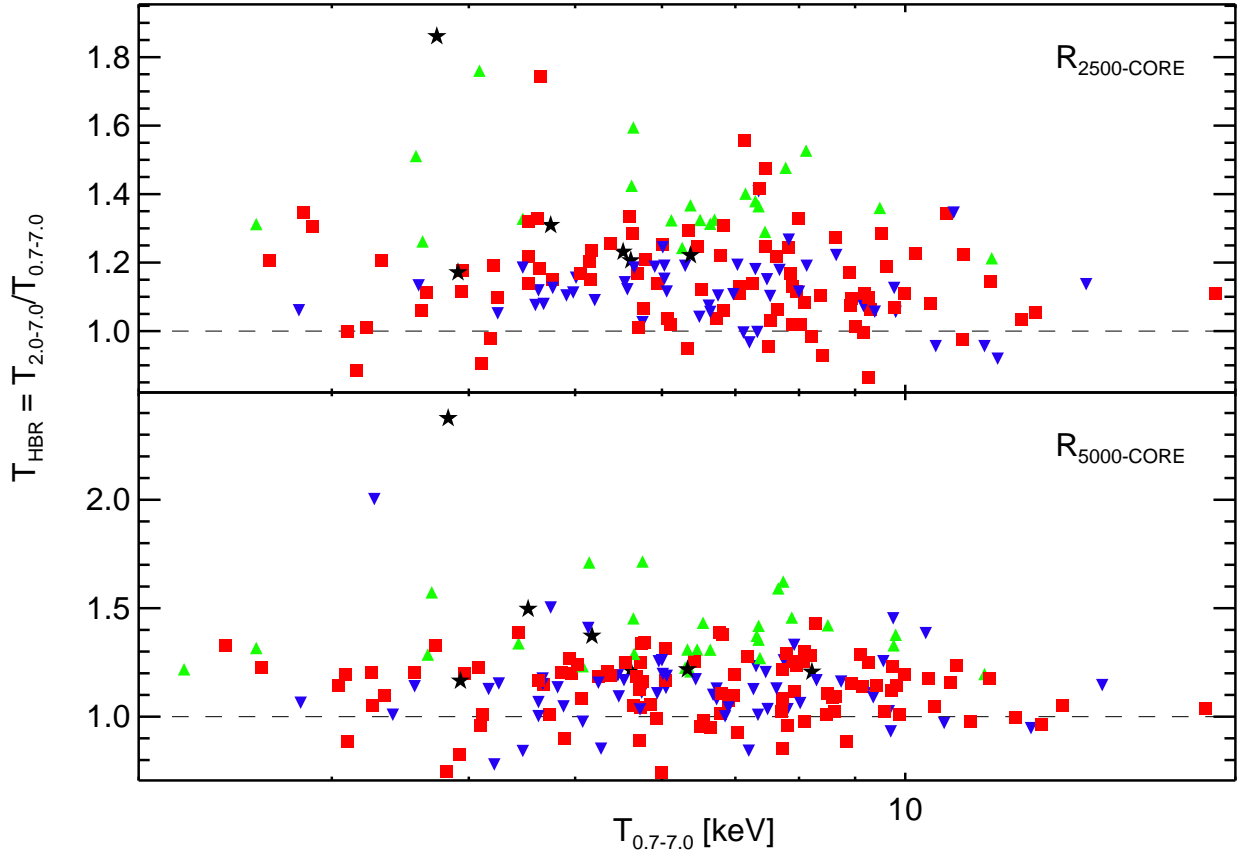}
\caption{
$T_{HBR}$ plotted against $T_{0.7-7.0}$ for the
$R_{2500-\mathrm{CORE}}$ and $R_{5000-\mathrm{CORE}}$ apertures. Note
the vertical scales for both panels are not the same. The top and
bottom panes contain 166 and 192 clusters respectively. Only two
clusters -- Abell 697 and MACS J2049.9-3217 -- do not have a $T_{HBR}
> 1.1$ in one aperture and not the other. In both cases though, it was
a result of narrowly missing the cut. The dashed lines are the lines
of equivalence. Symbols and color coding are based on two criteria: 1)
presence of a cool core (CC) and 2) value of $T_{HBR}$. Black stars (6
top, 7 bottom) are clusters with a CC and $T_{HBR}$ significantly
greater than 1.1. Green upright-triangles (21 top, 27 bottom) are NCC
clusters with $T_{HBR}$ significantly greater than 1.1. Blue
down-facing triangles (49 top, 60 bottom) are CC clusters and red
squares (90 top, 98 bottom) are NCC clusters. We have found most, if
not all, of the clusters with $T_{HBR} \gtrsim 1.1$ are merger
systems. Note that the cut at $T_{HBR} > 1.1$ is arbitrary and there
are more merger systems in our sample then just those highlighted in
this figure. However it is rather suggestive that clusters with the
highest values of $T_{HBR}$ appear to be merging systems.
}
\label{fig:ftx_tx}
\end{center}
\end{figure}
\clearpage

\LongTables
\clearpage
\begin{deluxetable}{lcccccccc}
\tablewidth{0pt}
\tabletypesize{\scriptsize}
\tablecaption{Summary of Sample\label{tab:sample}}
\tablehead{\colhead{Cluster} & \colhead{Obs.ID} & \colhead{R.A.} & \colhead{Dec.} & \colhead{ExpT} & \colhead{Mode} & \colhead{ACIS} & \colhead{$z$} & \colhead{$L_{bol.}$}\\
\colhead{ } & \colhead{ } & \colhead{hr:min:sec} & \colhead{$\degr:\arcmin:\arcsec$} & \colhead{ksec} & \colhead{ } & \colhead{ } & \colhead{ } & \colhead{$10^{44}$ ergs s$^{-1}$}\\
\colhead{{(1)}} & \colhead{{(2)}} & \colhead{{(3)}} & \colhead{{(4)}} & \colhead{{(5)}} & \colhead{{(6)}} & \colhead{{(7)}} & \colhead{{(8)}} & \colhead{{(9)}}
}
\startdata
1E0657 56 & \dataset [ADS/Sa.CXO\#obs/03184] {3184} & 06:58:29.622 & -55:56:39.79 & 87.5 & VF & I3 & 0.296 & 52.48\\
1E0657 56 & \dataset [ADS/Sa.CXO\#obs/05356] {5356} & 06:58:29.619 & -55:56:39.78 & 97.2 & VF & I2 & 0.296 & 52.48\\
1E0657 56 & \dataset [ADS/Sa.CXO\#obs/05361] {5361} & 06:58:29.620 & -55:56:39.80 & 82.6 & VF & I3 & 0.296 & 52.48\\
1RXS J2129.4-0741 & \dataset [ADS/Sa.CXO\#obs/03199] {3199} & 21:29:26.274 & -07:41:29.38 & 19.9 & VF & I3 & 0.570 & 20.58\\
1RXS J2129.4-0741 & \dataset [ADS/Sa.CXO\#obs/03595] {3595} & 21:29:26.281 & -07:41:29.36 & 19.9 & VF & I3 & 0.570 & 20.58\\
2PIGG J0011.5-2850 & \dataset [ADS/Sa.CXO\#obs/05797] {5797} & 00:11:21.623 & -28:51:14.44 & 19.9 & VF & I3 & 0.075 &  2.15\\
2PIGG J0311.8-2655 $\ddagger$ & \dataset [ADS/Sa.CXO\#obs/05799] {5799} & 03:11:33.904 & -26:54:16.48 & 39.6 & VF & I3 & 0.062 &  0.25\\
2PIGG J2227.0-3041 & \dataset [ADS/Sa.CXO\#obs/05798] {5798} & 22:27:54.560 & -30:34:34.84 & 22.3 & VF & I2 & 0.073 &  0.81\\
\enddata
\tablecomments{(1) Cluster name, (2) CDA observation identification
number, (3) R.A. of cluster center, (4) Dec. of cluster center, (5)
nominal exposure time, (6) observing mode, (7) CCD location of
centroid, (8) redshift, (9) bolometric luminosity. A ($\ddagger$)
indicates a cluster analyzed within R$_{5000}$ only. Italicized
cluster names indicate a cluster which was excluded from our analysis
(discussed in \S\ref{sec:fitting}). For clusters with multiple
observations, the X-ray centers differ by $< 0.5$ kpc.}
\end{deluxetable}
\begin{deluxetable}{cc|cc}
\tabletypesize{\scriptsize}
\tablecaption{Summary of two-component simulations\label{tab:simres}}
\tablewidth{0pt}
\tablehead{
\colhead{$T_2$} & \colhead{$\xi_{min}$} & \colhead{$T_2$} & \colhead{$\xi_{min}$}\\
\colhead{keV}   & \colhead{--} & \colhead{keV} & \colhead{--}\\
\multicolumn{2}{c|}{Idealized Spectra} & \multicolumn{2}{c}{Observation-Specific Spectra}}
\startdata
0.50 & $\geq 12\% \pm 4\%$ & 0.50 & $\geq 14.5\% \pm 0.1\%$\\
0.75 & $\geq 12\% \pm 4\%$ & 0.75 & $\geq 11.7\% \pm 0.1\%$\\
1.00 & $\geq 8\% \pm 3\%$  & 1.00 & $\geq 11.6\% \pm 0.1\%$\\
1.25 & $\geq 17\% \pm 3\%$ & \dotfill & \dotfill\\
1.50 & $\geq 23\% \pm 5\%$ & \dotfill & \dotfill\\
1.75 & $\geq 28\% \pm 4\%$ & \dotfill & \dotfill\\
2.00 & none                & 2.00 & $\geq 25.5\% \pm 0.1\%$\\
3.00 & none                & 3.00 & $\geq 28.9\% \pm 0.1\%$
\enddata
\tablecomments{
This table summarizes the results of the two temperature component
spectra simulations for the ideal and observation-specific cases (see
\S\ref{sec:simulated} for details). The parameter $\xi_{min}$
represents the minimum fractional contribution of the cooler
component, $T_2$, to the overall count rate in order to produce
$T_{HBR} \geq 1.1$ at 90\% confidence. The results for the
observation-specific spectra are for spectra with $N_{\mathrm{counts}}
> 15,000$.
}
\end{deluxetable}

\begin{deluxetable}{c|ccc|ccc}
\tabletypesize{\scriptsize}
\tablecaption{Weighted averages for various apertures\label{tab:wavg}}
\tablewidth{0pt}
\tablehead{
\colhead{ } & 
\colhead{[0.7-7.0]} & \colhead{[2.0-7.0]} & \colhead{$T_{HBR}$} & 
\colhead{[0.7-7.0]} & \colhead{[2.0-7.0]} & \colhead{$T_{HBR}$}\\
\colhead{ } & \colhead{keV} & \colhead{keV} &
\colhead{ } & \colhead{keV} & \colhead{keV} & \colhead{ }\\
\multicolumn{1}{c|}{Aperture} & \multicolumn{3}{l}{\dotfill Without
Core\dotfill} & \multicolumn{3}{l}{\dotfill With Core\dotfill}}
\startdata
R$_{2500}$ & 4.93$\pm 0.03$   & 6.24$\pm 0.07$   & 1.16$\pm 0.01$   & 4.47$\pm 0.02$ & 5.45$\pm 0.05$ & 1.13$\pm 0.01$\\ 
R$_{5000}$ & 4.75$\pm 0.02$   & 5.97$\pm 0.07$   & 1.14$\pm 0.01$   & 4.27$\pm 0.02$ & 5.29$\pm 0.05$ & 1.14$\pm 0.01$\\
Simulated  & 3.853$\pm 0.004$ & 4.457$\pm 0.009$ & 1.131$\pm 0.002$ & \dotfill       & \dotfill       & \dotfill\\
Control    & 4.208$\pm 0.003$ & 4.468$\pm 0.006$ & 1.002$\pm 0.001$ & \dotfill       & \dotfill       & \dotfill
\enddata
\tablecomments{
Quoted errors are standard deviation of the mean calculated using an
unbiased estimator for weighted samples. Simulated sample has been
culled to include only T$_2$=0.75 keV.
}
\end{deluxetable}
\begin{deluxetable}{lcccclc}
\tabletypesize{\scriptsize}
\tablecaption{Clusters with $T_{HBR}$ $> 1.1$ with 90\% confidence.\label{tab:tf11}}
\tablewidth{0pt}
\tablehead{
\colhead{Name}       & \colhead{$T_{HBR}$} & \colhead{Merger?} &
\colhead{Core Class} & \colhead{$T_{dec}$} & \colhead{X-ray Morphology} & 
\colhead{Ref.}}
\startdata
RX J1525+0958       \dotfill & 1.86$^{+0.83}_{-0.51}$ & Y       &  CC & 0.42$^{+0.14}_{-0.08}$ & Arrowhead shape \& no discernible core & [29]\\
MS 1008.1-1224      \dotfill & 1.59$^{+0.37}_{-0.27}$ & Y       & NCC & 0.93$^{+0.19}_{-0.14}$ & Wide gas tail extending $\approx$550 kpc north & [1]\\
ABELL 2034          \dotfill & 1.40$^{+0.14}_{-0.11}$ & Y       & NCC & 1.07$^{+0.11}_{-0.09}$ & Prominent cold front \& gas tail extending south & [2]\\
ABELL 401           \dotfill & 1.37$^{+0.12}_{-0.10}$ & Y       & NCC & 1.13$^{+0.12}_{-0.10}$ & Highly spherical \& possible cold front to north & [3]\\
ABELL 1689          \dotfill & 1.36$^{+0.14}_{-0.12}$ & Y       & NCC & 0.95$^{+0.09}_{-0.07}$ & Exceptionally spherical \& bright central core & [6],[7]\\
RX J0439.0+0715     \dotfill & 1.42$^{+0.24}_{-0.18}$ & Unknown & NCC & 0.98$^{+0.11}_{-0.09}$ & Bright core \& possible cold front to north & [29]\\
ABELL 3376          \dotfill & 1.33$^{+0.11}_{-0.10}$ & Y       & NCC & 0.97$^{+0.07}_{-0.07}$ & Highly disturbed \& broad gas tail to west & [4],[5]\\
ABELL 2255          \dotfill & 1.32$^{+0.12}_{-0.10}$ & Y       & NCC & 1.48$^{+0.32}_{-0.23}$ & Spherical \& compressed isophotes west of core & [8],[9]\\
\enddata
\tablecomments{Clusters ordered by lower limit of $T_{HBR}$. Listed
$T_{HBR}$ values are for the $R_{2500-\mathrm{CORE}}$ aperture, with
the exception of the ``$R_{5000-\mathrm{CORE}}$ Only'' clusters listed
at the end of the table. Excluding the ``$R_{5000-\mathrm{CORE}}$
Only'' clusters, all clusters listed here had $T_{HBR}$ significantly
greater than 1.1 and the same core classification for both the
$R_{2500-\mathrm{CORE}}$ and $R_{5000-\mathrm{CORE}}$ apertures.
[1]  \cite{1994ApJS...94..583G}, 
[2]  \cite{2003ApJ...593..291K}, 
[3]  \cite{2005ChJAA...5..126Y}, 
[4]  \cite{1998ApJ...503...77M}, 
[5]  \cite{2006Sci...314..791B}, 
[6]  \cite{1990ApJS...72..715T}, 
[7]  \cite{2004ApJ...607..190A}, 
[8]  \cite{1995ApJ...446..583B}, 
[9]  \cite{1997A&A...317..432F}, 
[10] \cite{1997ApJ...490...56G}, 
[11] \cite{2002ApJS..139..313D}, 
[12] \cite{2005MNRAS.359..417S}, 
[13] \cite{1982ApJ...255L..17G}, 
[14] \cite{2004ApJ...610L..81H}, 
[15] \cite{2004ApJ...614..692Y}, 
[16] \cite{2003A&A...408...57M}, 
[17] \cite{2000ApJ...540..726G}, 
[18] \cite{1998ApJ...496L...5T}, 
[19] \cite{1999AcA....49..403K}, 
[20] \cite{2001ApJ...555..205M}, 
[21] \cite{2001A&A...379..807G}, 
[22] \cite{1998MNRAS.301..609B}, 
[23] \cite{2005ApJ...619..161G}, 
[24] \cite{1996ApJ...472L..17M}, 
[25] \cite{2001ASPC..251..474O}, 
[26] \cite{2000ApJ...534L..43M}, 
[27] \cite{2004ApJ...616..178C}, 
[28] \cite{2005xrrc.procE7.08C}, 
[29] this work.
}
\end{deluxetable}

\begin{deluxetable}{lcccccccccc}
\tablewidth{0pt}
\tabletypesize{\scriptsize}
\tablecaption{Summary of Excised R$_{2500}$ Spectral Fits\label{tab:r2500specfits}}
\tablehead{\colhead{Cluster} & \colhead{R$_{\mathrm{CORE}}$} & \colhead{R$_{2500}$ } & \colhead{N$_{HI}$} & \colhead{T$_{77}$} & \colhead{T$_{27}$} & \colhead{T$_{HBR}$} & \colhead{Z$_{77}$} & \colhead{$\chi^2_{red,77}$} & \colhead{$\chi^2_{red,27}$} & \colhead{\% Source}\\
\colhead{ } & \colhead{kpc} & \colhead{kpc} & \colhead{$10^{20}$ cm$^{-2}$} & \colhead{keV} & \colhead{keV} & \colhead{ } & \colhead{Z$_{\sun}$} & \colhead{ } & \colhead{ } & \colhead{ }\\
\colhead{{(1)}} & \colhead{{(2)}} & \colhead{{(3)}} & \colhead{{(4)}} & \colhead{{(5)}} & \colhead{{(6)}} & \colhead{{(7)}} & \colhead{{(8)}} & \colhead{{(9)}} & \colhead{{(10)}} & \colhead{{(11)}}
}
\startdata
1E0657 56 $\star$ &    69 &   688 & 6.53  & 11.99  $^{+0.27   }_{-0.26   }$  & 14.54  $^{+0.67   }_{-0.53   }$  & 1.21   $^{+0.06   }_{-0.05   }$  & 0.29$^{+0.03   }_{-0.02   }$  & 1.24 & 1.11 &  92\\
1RXS J2129.4-0741 $\star$ &    71 &   526 & 4.36  & 8.22   $^{+1.18   }_{-0.95   }$  & 8.10   $^{+1.47   }_{-1.10   }$  & 0.99   $^{+0.23   }_{-0.18   }$  & 0.43$^{+0.18   }_{-0.17   }$  & 1.07 & 1.05 &  80\\
2PIGG J0011.5-2850 &    69 &   547 & 2.18  & 5.15   $^{+0.25   }_{-0.24   }$  & 6.20   $^{+0.79   }_{-0.65   }$  & 1.20   $^{+0.16   }_{-0.14   }$  & 0.26$^{+0.09   }_{-0.08   }$  & 1.09 & 1.00 &  70\\
2PIGG J2227.0-3041 &    69 &   378 & 1.11  & 2.80   $^{+0.15   }_{-0.14   }$  & 2.97   $^{+0.34   }_{-0.27   }$  & 1.06   $^{+0.13   }_{-0.11   }$  & 0.35$^{+0.09   }_{-0.08   }$  & 1.16 & 1.15 &  69\\
3C 220.1 &    71 &   456 & 1.91  & 9.26   $^{+14.71  }_{-3.98   }$  & 8.00   $^{+17.66  }_{-4.03   }$  & 0.86   $^{+2.35   }_{-0.57   }$  & 0.00$^{+0.59   }_{-0.00   }$  & 1.20 & 1.40 &  30\\
3C 28.0 &    70 &   420 & 5.71  & 5.53   $^{+0.29   }_{-0.27   }$  & 6.81   $^{+0.71   }_{-0.60   }$  & 1.23   $^{+0.14   }_{-0.12   }$  & 0.30$^{+0.08   }_{-0.07   }$  & 0.98 & 0.88 &  87\\
3C 295 &    69 &   465 & 1.35  & 5.16   $^{+0.42   }_{-0.38   }$  & 5.93   $^{+0.84   }_{-0.69   }$  & 1.15   $^{+0.19   }_{-0.16   }$  & 0.38$^{+0.12   }_{-0.11   }$  & 0.91 & 0.93 &  79\\
3C 388 &    69 &   420 & 6.11  & 3.23   $^{+0.23   }_{-0.21   }$  & 3.26   $^{+0.49   }_{-0.37   }$  & 1.01   $^{+0.17   }_{-0.13   }$  & 0.51$^{+0.16   }_{-0.14   }$  & 0.95 & 0.95 &  68\\
\enddata
\tablecomments{Note: ``77'' refers to 0.7-7.0 keV band and ``27''
refers to 2.0-7.0 keV band. (1) Cluster name, (2) size of excluded
core region in kpc, (3) $R_{2500}$ in kpc, (4) absorbing Galactic
neutral hydrogen column density, (5,6) best-fit {\textsc{MeKaL}}
temperatures, (7) $T_{0.7-7.0}$/$T_{2.0-7.0}$ also called $T_{HBR}$,
(8) best-fit 77 {\textsc{MeKaL}} abundance, (9,10) respective
reduced $\chi^2$ statistics, and (11) percent of emission attributable
to source. A star ($\star$) indicates a cluster which has multiple
observations. Each observation has an independent spectrum extracted
along with an associated WARF, WRMF, normalized background spectrum,
and soft residual. Each independent spectrum is then fit
simultaneously with the same spectral model to produce the final fit.}
\end{deluxetable}

\begin{deluxetable}{lcccccccccc}
\tablewidth{0pt}
\tabletypesize{\scriptsize}
\tablecaption{Summary of Excised R$_{5000}$ Spectral Fits\label{tab:r5000specfits}}
\tablehead{\colhead{Cluster} & \colhead{R$_{\mathrm{CORE}}$} & \colhead{R$_{5000}$ } & \colhead{N$_{HI}$} & \colhead{T$_{77}$} & \colhead{T$_{27}$} & \colhead{T$_{HBR}$} & \colhead{Z$_{77}$} & \colhead{$\chi^2_{red,77}$} & \colhead{$\chi^2_{red,27}$} & \colhead{\% Source}\\
\colhead{ } & \colhead{kpc} & \colhead{kpc} & \colhead{$10^{20}$ cm$^{-2}$} & \colhead{keV} & \colhead{keV} & \colhead{ } & \colhead{Z$_{\sun}$} & \colhead{ } & \colhead{ } & \colhead{ }\\
\colhead{{(1)}} & \colhead{{(2)}} & \colhead{{(3)}} & \colhead{{(4)}} & \colhead{{(5)}} & \colhead{{(6)}} & \colhead{{(7)}} & \colhead{{(8)}} & \colhead{{(9)}} & \colhead{{(10)}} & \colhead{{(11)}}
}
\startdata
1E0657 56 $\star$ &    69 &   487 & 6.53  & 11.81  $^{+0.29   }_{-0.27   }$  & 14.13  $^{+0.58   }_{-0.53   }$  & 1.20   $^{+0.06   }_{-0.05   }$  & 0.29$^{+0.03   }_{-0.03   }$  & 1.22 & 1.10 &  95\\
1RXS J2129.4-0741 $\star$ &    71 &   373 & 4.36  & 8.47   $^{+1.31   }_{-1.04   }$  & 8.57   $^{+1.73   }_{-1.27   }$  & 1.01   $^{+0.26   }_{-0.19   }$  & 0.51$^{+0.20   }_{-0.19   }$  & 1.16 & 1.27 &  87\\
2PIGG J0011.5-2850 &    69 &   387 & 2.18  & 5.25   $^{+0.29   }_{-0.27   }$  & 6.21   $^{+0.83   }_{-0.68   }$  & 1.18   $^{+0.17   }_{-0.14   }$  & 0.23$^{+0.09   }_{-0.08   }$  & 1.08 & 1.01 &  78\\
2PIGG J0311.8-2655 &    69 &   321 & 1.46  & 3.35   $^{+0.25   }_{-0.22   }$  & 3.67   $^{+0.71   }_{-0.54   }$  & 1.10   $^{+0.23   }_{-0.18   }$  & 0.33$^{+0.13   }_{-0.11   }$  & 1.03 & 1.10 &  51\\
2PIGG J2227.0-3041 &    69 &   267 & 1.11  & 2.81   $^{+0.16   }_{-0.15   }$  & 2.99   $^{+0.36   }_{-0.28   }$  & 1.06   $^{+0.14   }_{-0.11   }$  & 0.35$^{+0.11   }_{-0.08   }$  & 1.14 & 1.10 &  77\\
3C 220.1 &    71 &   322 & 1.91  & 7.81   $^{+7.50   }_{-2.99   }$  & 7.49   $^{+11.53  }_{-3.51   }$  & 0.96   $^{+1.74   }_{-0.58   }$  & 0.00$^{+0.55   }_{-0.00   }$  & 0.60 & 0.78 &  36\\
3C 28.0 &    70 &   297 & 5.71  & 5.18   $^{+0.28   }_{-0.27   }$  & 7.11   $^{+1.15   }_{-0.90   }$  & 1.37   $^{+0.23   }_{-0.19   }$  & 0.30$^{+0.09   }_{-0.07   }$  & 0.96 & 0.77 &  90\\
3C 295 &    69 &   329 & 1.35  & 5.47   $^{+0.49   }_{-0.42   }$  & 6.51   $^{+0.92   }_{-0.78   }$  & 1.19   $^{+0.20   }_{-0.17   }$  & 0.29$^{+0.11   }_{-0.11   }$  & 1.02 & 1.04 &  87\\
\enddata
\tablecomments{Note: ``77'' refers to 0.7-7.0 keV band and ``27''
refers to 2.0-7.0 keV band. (1) Cluster name, (2) size of excluded
core region in kpc, (3) $R_{5000}$ in kpc, (4) absorbing Galactic
neutral hydrogen column density, (5,6) best-fit {\textsc{MeKaL}}
temperatures, (7) $T_{0.7-7.0}$/$T_{2.0-7.0}$ also called $T_{HBR}$,
(8) best-fit 77 {\textsc{MeKaL}} abundance, (9,10) respective
reduced $\chi^2$ statistics, and (11) percent of emission attributable
to source. A star ($\star$) indicates a cluster which has multiple
observations. Each observation has an independent spectrum extracted
along with an associated WARF, WRMF, normalized background spectrum,
and soft residual. Each independent spectrum is then fit
simultaneously with the same spectral model to produce the final fit.}
\end{deluxetable}

\begin{thebibliography}{} 

\bibitem[{{Akritas} \& {Bershady}(1996)}]{1996ApJ...470..706A}
{Akritas}, M.~G. \& {Bershady}, M.~A. 1996, \apj, 470, 706

\bibitem[{{Anders} \& {Grevesse}(1989)}]{1989GeCoA..53..197A}
{Anders}, E., \& {Grevesse}, N. 1989, \gca, 53, 197

\bibitem[{{Andersson} \& {Madejski}(2004)}]{2004ApJ...607..190A}
{Andersson}, K.~E., \& {Madejski}, G.~M. 2004, \apj, 607, 190

\bibitem[{{Arnaud}(1996)}]{1996ASPC..101...17A}
{Arnaud}, K.~A. 1996, in ASP Conf. Ser. 101: Astronomical Data Analysis
  Software and Systems V, ed. G.~H. {Jacoby} \& J.~{Barnes}, 17--+

\bibitem[{{Arnaud} {et~al.}(2002){Arnaud}, {Aghanim}, \&
  {Neumann}}]{2002A&A...389....1A}
{Arnaud}, M., {Aghanim}, N., \& {Neumann}, D.~M. 2002, \aap, 389, 1

\bibitem[{{Bagchi} {et~al.}(2006){Bagchi}, {Durret}, {Neto}, \&
  {Paul}}]{2006Sci...314..791B}
{Bagchi}, J., {Durret}, F., {Neto}, G.~B.~L., \& {Paul}, S. 2006, Science, 314,
  791

\bibitem[{{Balestra} {et~al.}(2007){Balestra}, {Tozzi}, {Ettori}, {Rosati},
  {Borgani}, {Mainieri}, {Norman}, \& {Viola}}]{2007A&A...462..429B}
{Balestra}, I., {Tozzi}, P., {Ettori}, S., {Rosati}, P., {Borgani}, S.,
  {Mainieri}, V., {Norman}, C., \& {Viola}, M. 2007, \aap, 462, 429

\bibitem[{{Barrena} {et~al.}(2007){Barrena}, {Boschin}, {Girardi}, \&
  {Spolaor}}]{2007A&A...467...37B}
{Barrena}, R., {Boschin}, W., {Girardi}, M., \& {Spolaor}, M. 2007, \aap, 467,
  37

\bibitem[{{Bauer} {et~al.}(2005){Bauer}, {Fabian}, {Sanders}, {Allen}, \&
  {Johnstone}}]{2005MNRAS.359.1481B}
{Bauer}, F.~E., {Fabian}, A.~C., {Sanders}, J.~S., {Allen}, S.~W., \&
  {Johnstone}, R.~M. 2005, \mnras, 359, 1481

\bibitem[{{Bliton} {et~al.}(1998){Bliton}, {Rizza}, {Burns}, {Owen}, \&
  {Ledlow}}]{1998MNRAS.301..609B}
{Bliton}, M., {Rizza}, E., {Burns}, J.~O., {Owen}, F.~N., \& {Ledlow}, M.~J.
  1998, \mnras, 301, 609

\bibitem[{{B{\"o}hringer} {et~al.}(2007){B{\"o}hringer}, {Schuecker},
  {Pratt}, {Arnaud}, {Ponman}, {Croston}, {Borgani}, {Bower}, {Briel},
  {Collins}, {Donahue}, {Forman}, {Finoguenov}, {Geller}, {Guzzo},
  {Henry}, {Kneissl}, {Mohr}, {Matsushita}, {Mullis}, {Ohashi},
  {Pedersen}, {Pierini}, {Quintana}, {Raychaudhury}, {Reiprich},
  {Romer}, {Rosati}, {Sabirli}, {Temple}, {Viana}, {Vikhlinin}, {Voit},
  {Zhang}}]{2007A&A...469..363B}
{B{\"o}hringer}, H., {Schuecker}, P., {Pratt}, G.~W., {Arnaud}, M.,
  {Ponman}, T.~J., {Croston}, J.~H., {Borgani}, S., {Bower}, R.~G.,
  {Briel}, U.~G., {Collins}, C.~A., {Donahue}, M., {Forman}, W.~R.,
  {Finoguenov}, A., {Geller}, M.~J., {Guzzo}, L., {Henry}, J.~P.,
  {Kneissl}, R., {Mohr}, J.~J., {Matsushita}, K., {Mullis}, C.~R.,
  {Ohashi}, T., {Pedersen}, K., {Pierini}, D., {Quintana}, H.,
  {Raychaudhury}, S., {Reiprich}, T.~H., {Romer}, A.~K., {Rosati}, P.,
  {Sabirli}, K., {Temple}, R.~F., {Viana}, P.~T.~P., {Vikhlinin}, A.,
  {Voit}, G.~M., \& {Zhang}, Y.-Y. 2007, \aap, 469, 363

\bibitem[{{B{\"o}hringer} {et~al.}(2004){B{\"o}hringer}, {Schuecker}, {Guzzo},
  {Collins}, {Voges}, {Cruddace}, {Ortiz-Gil}, {Chincarini}, {De Grandi},
  {Edge}, {MacGillivray}, {Neumann}, {Schindler}, \&
  {Shaver}}]{2004A&A...425..367B}
{B{\"o}hringer}, H., {Schuecker}, P., {Guzzo}, L., {Collins}, C.~A., {Voges},
  W., {Cruddace}, R.~G., {Ortiz-Gil}, A., {Chincarini}, G., {De Grandi}, S.,
  {Edge}, A.~C., {MacGillivray}, H.~T., {Neumann}, D.~M., {Schindler}, S., \&
  {Shaver}, P. 2004, \aap, 425, 367

\bibitem[{{Buote} \& {Tsai}(1995)}]{1995ApJ...452..522B}
{Buote}, D.~A., \& {Tsai}, J.~C. 1995, \apj, 452, 522

\bibitem[{{Buote} \& {Tsai}(1996)}]{1996ApJ...458...27B}
---. 1996, \apj, 458, 27

\bibitem[{{Burns} {et~al.}(2007){Burns}, {Hallman}, {Gantner}, {Motl}, \&
  {Norman}}]{2007arXiv0708.1954B}
{Burns}, J.~O., {Hallman}, E.~J., {Gantner}, B., {Motl}, P.~M., \& {Norman},
  M.~L. 2007, ArXiv e-prints, 708

\bibitem[{{Burns} {et~al.}(1995){Burns}, {Roettiger}, {Pinkney}, {Perley},
  {Owen}, \& {Voges}}]{1995ApJ...446..583B}
{Burns}, J.~O., {Roettiger}, K., {Pinkney}, J., {Perley}, R.~A., {Owen}, F.~N.,
  \& {Voges}, W. 1995, \apj, 446, 583

\bibitem[{{Cash}(1979)}]{1979ApJ...228..939C}
{Cash}, W. 1979, \apj, 228, 939

\bibitem[{{Chen} {et~al.}(2007){Chen}, {Reiprich}, {B{\"o}hringer}, {Ikebe}, \&
  {Zhang}}]{2007A&A...466..805C}
{Chen}, Y., {Reiprich}, T.~H., {B{\"o}hringer}, H., {Ikebe}, Y., \& {Zhang},
  Y.-Y. 2007, \aap, 466, 805

\bibitem[{{Clarke} {et~al.}(2004){Clarke}, {Blanton}, \&
  {Sarazin}}]{2004ApJ...616..178C} 
{Clarke}, T.~E., {Blanton}, E.~L., \& {Sarazin}, C.~L. 2004, \apj,
  616 178

\bibitem[{{Clarke} {et~al.}(2005){Clarke}, {Blanton}, \&
  {Sarazin}}]{2005xrrc.procE7.08C}
{Clarke}, T.~E., {Blanton}, E.~L., \& {Sarazin}, C.~L. 2005, X-Ray
  and Radio Connections, ed. 7

\bibitem[{{Dahle} {et~al.}(2002){Dahle}, {Kaiser}, {Irgens}, {Lilje}, \&
  {Maddox}}]{2002ApJS..139..313D}
{Dahle}, H., {Kaiser}, N., {Irgens}, R.~J., {Lilje}, P.~B., \& {Maddox}, S.~J.
  2002, \apjs, 139, 313

\bibitem[{{David} \& {Kempner}(2004)}]{2004ApJ...613..831D}
{David}, L.~P., \& {Kempner}, J. 2004, \apj, 613, 831

\bibitem[{{Dickey} \& {Lockman}(1990)}]{1990ARA&A..28..215D}
{Dickey}, J.~M., \& {Lockman}, F.~J. 1990, \araa, 28, 215

\bibitem[{{Ebeling} {et~al.}(2000){Ebeling}, {Edge}, {Allen}, {Crawford},
  {Fabian}, \& {Huchra}}]{2000MNRAS.318..333E}
{Ebeling}, H., {Edge}, A.~C., {Allen}, S.~W., {Crawford}, C.~S., {Fabian},
  A.~C., \& {Huchra}, J.~P. 2000, \mnras, 318, 333

\bibitem[{{Ebeling} {et~al.}(1998){Ebeling}, {Edge}, {Bohringer}, {Allen},
  {Crawford}, {Fabian}, {Voges}, \& {Huchra}}]{1998MNRAS.301..881E}
{Ebeling}, H., {Edge}, A.~C., {Bohringer}, H., {Allen}, S.~W., {Crawford},
  C.~S., {Fabian}, A.~C., {Voges}, W., \& {Huchra}, J.~P. 1998, \mnras, 301,
  881

\bibitem[{{Ebeling} {et~al.}(2001){Ebeling}, {Edge}, \&
  {Henry}}]{2001ApJ...553..668E}
{Ebeling}, H., {Edge}, A.~C., \& {Henry}, J.~P. 2001, \apj, 553, 668

\bibitem[{{Edge} {et~al.}(1990){Edge}, {Stewart}, {Fabian}, \&
  {Arnaud}}]{1990MNRAS.245..559E}
{Edge}, A.~C., {Stewart}, G.~C., {Fabian}, A.~C., \& {Arnaud}, K.~A. 1990,
  \mnras, 245, 559

\bibitem[{{Eisenstein} {et~al.}(2005){Eisenstein}, {Zehavi}, {Hogg},
  {Scoccimarro}, {Blanton}, {Nichol}, {Scranton}, {Seo}, {Tegmark}, {Zheng},
  {Anderson}, {Annis}, {Bahcall}, {Brinkmann}, {Burles}, {Castander},
  {Connolly}, {Csabai}, {Doi}, {Fukugita}, {Frieman}, {Glazebrook}, {Gunn},
  {Hendry}, {Hennessy}, {Ivezi{\'c}}, {Kent}, {Knapp}, {Lin}, {Loh}, {Lupton},
  {Margon}, {McKay}, {Meiksin}, {Munn}, {Pope}, {Richmond}, {Schlegel},
  {Schneider}, {Shimasaku}, {Stoughton}, {Strauss}, {SubbaRao}, {Szalay},
  {Szapudi}, {Tucker}, {Yanny}, \& {York}}]{2005ApJ...633..560E}
{Eisenstein}, D.~J., {Zehavi}, I., {Hogg}, D.~W., {Scoccimarro}, R., {Blanton},
  M.~R., {Nichol}, R.~C., {Scranton}, R., {Seo}, H.-J., {Tegmark}, M., {Zheng},
  Z., {Anderson}, S.~F., {Annis}, J., {Bahcall}, N., {Brinkmann}, J., {Burles},
  S., {Castander}, F.~J., {Connolly}, A., {Csabai}, I., {Doi}, M., {Fukugita},
  M., {Frieman}, J.~A., {Glazebrook}, K., {Gunn}, J.~E., {Hendry}, J.~S.,
  {Hennessy}, G., {Ivezi{\'c}}, Z., {Kent}, S., {Knapp}, G.~R., {Lin}, H.,
  {Loh}, Y.-S., {Lupton}, R.~H., {Margon}, B., {McKay}, T.~A., {Meiksin}, A.,
  {Munn}, J.~A., {Pope}, A., {Richmond}, M.~W., {Schlegel}, D., {Schneider},
  D.~P., {Shimasaku}, K., {Stoughton}, C., {Strauss}, M.~A., {SubbaRao}, M.,
  {Szalay}, A.~S., {Szapudi}, I., {Tucker}, D.~L., {Yanny}, B., \& {York},
  D.~G. 2005, \apj, 633, 560

\bibitem[{{Evrard}(1989)}]{1989ApJ...341L..71E}
{Evrard}, A.~E. 1989, \apjl, 341, L71

\bibitem[{{Evrard} {et~al.}(1996){Evrard}, {Metzler}, \&
  {Navarro}}]{1996ApJ...469..494E}
{Evrard}, A.~E., {Metzler}, C.~A., \& {Navarro}, J.~F. 1996, \apj, 469, 494

\bibitem[{{Feretti} {et~al.}(1997){Feretti}, {Boehringer}, {Giovannini}, \&
  {Neumann}}]{1997A&A...317..432F}
{Feretti}, L., {Boehringer}, H., {Giovannini}, G., \& {Neumann}, D. 1997, \aap,
  317, 432

\bibitem[{{Freeman} {et~al.}(2002){Freeman}, {Kashyap}, {Rosner}, \&
  {Lamb}}]{2002ApJS..138..185F}
{Freeman}, P.~E., {Kashyap}, V., {Rosner}, R., \& {Lamb}, D.~Q. 2002, \apjs,
  138, 185

\bibitem[{{Gioia} {et~al.}(1982){Gioia}, {Maccacaro}, {Geller}, {Huchra},
  {Stocke}, \& {Steiner}}]{1982ApJ...255L..17G}
{Gioia}, I.~M., {Maccacaro}, T., {Geller}, M.~J., {Huchra}, J.~P., {Stocke},
  J., \& {Steiner}, J.~E. 1982, \apjl, 255, L17

\bibitem[{{Gioia} {et~al.}(1990){Gioia}, {Maccacaro}, {Schild}, {Wolter},
  {Stocke}, {Morris}, \& {Henry}}]{1990ApJS...72..567G}
{Gioia}, I.~M., {Maccacaro}, T., {Schild}, R.~E., {Wolter}, A., {Stocke},
  J.~T., {Morris}, S.~L., \& {Henry}, J.~P. 1990, \apjs, 72, 567

\bibitem[{{Gioia} \& {Luppino}(1994)}]{1994ApJS...94..583G}
{Gioia}, I.~M. \& {Luppino}, G.~A. 1994, \apjs, 94, 583

\bibitem[{{Girardi} {et~al.}(1997){Girardi}, {Fadda}, {Escalera}, {Giuricin},
  {Mardirossian}, \& {Mezzetti}}]{1997ApJ...490...56G}
{Girardi}, M., {Fadda}, D., {Escalera}, E., {Giuricin}, G., {Mardirossian}, F.,
  \& {Mezzetti}, M. 1997, \apj, 490, 56

\bibitem[{{G{\'o}mez} {et~al.}(2000){G{\'o}mez}, {Hughes}, \&
  {Birkinshaw}}]{2000ApJ...540..726G}
{G{\'o}mez}, P.~L., {Hughes}, J.~P., \& {Birkinshaw}, M. 2000, \apj, 540, 726

\bibitem[{{Govoni} {et~al.}(2001){Govoni}, {Taylor}, {Dallacasa}, {Feretti}, \&
  {Giovannini}}]{2001A&A...379..807G}
{Govoni}, F., {Taylor}, G.~B., {Dallacasa}, D., {Feretti}, L., \& {Giovannini},
  G. 2001, \aap, 379, 807

\bibitem[{{Gutierrez} \& {Krawczynski}(2005)}]{2005ApJ...619..161G}
{Gutierrez}, K., \& {Krawczynski}, H. 2005, \apj, 619, 161

\bibitem[{{Haiman} {et~al.}(2001){Haiman}, {Mohr}, \&
  {Holder}}]{2001ApJ...553..545H}
{Haiman}, Z., {Mohr}, J.~J., \& {Holder}, G.~P. 2001, \apj, 553, 545

\bibitem[{{Hallman} \& {Markevitch}(2004)}]{2004ApJ...610L..81H}
{Hallman}, E.~J., \& {Markevitch}, M. 2004, \apjl, 610, L81

\bibitem[{{Henry} {et~al.}(2006){Henry}, {Mullis}, {Voges}, {B{\"o}hringer},
  {Briel}, {Gioia}, \& {Huchra}}]{2006ApJS..162..304H}
{Henry}, J.~P., {Mullis}, C.~R., {Voges}, W., {B{\"o}hringer}, H., {Briel},
  U.~G., {Gioia}, I.~M., \& {Huchra}, J.~P. 2006, \apjs, 162, 304

\bibitem[{{Jeltema} {et~al.}(2005){Jeltema}, {Canizares}, {Bautz}, \&
  {Buote}}]{2005ApJ...624..606J}
{Jeltema}, T.~E., {Canizares}, C.~R., {Bautz}, M.~W., \& {Buote}, D.~A. 2005,
  \apj, 624, 606

\bibitem[{{Jeltema} {et~al.}(2007){Jeltema}, {Hallman}, {Burns}, \&
  {Motl}}]{2007arXiv0708.1518J}
{Jeltema}, T.~E., {Hallman}, E.~J., {Burns}, J.~O., \& {Motl}, P.~M. 2007,
  ArXiv e-prints, 708

\bibitem[{{Juett} {et~al.}(2008){Juett}, {Sarazin}, {Clarke}, {Andernach},
  {Ehle}, {Fujita}, {Kempner}, {Roy}, {Rudnick}, \&
  {Slee}}]{2008ApJ...672..138J}
{Juett}, A.~M., {Sarazin}, C.~L., {Clarke}, T.~E., {Andernach}, H., {Ehle}, M.,
  {Fujita}, Y., {Kempner}, J.~C., {Roy}, A.~L., {Rudnick}, L., \& {Slee}, O.~B.
  2008, \apj, 672, 138

\bibitem[{{Kaastra}(1992)}]{1992SRON}
{Kaastra}, J.~S. 1992

\bibitem[{{Kempner} {et~al.}(2003){Kempner}, {Sarazin}, \&
  {Markevitch}}]{2003ApJ...593..291K}
{Kempner}, J.~C., {Sarazin}, C.~L., \& {Markevitch}, M. 2003, \apj, 593, 291

\bibitem[{{Kravtsov} {et~al.}(2006){Kravtsov}, {Vikhlinin}, \&
  {Nagai}}]{2006ApJ...650..128K}
{Kravtsov}, A.~V., {Vikhlinin}, A., \& {Nagai}, D. 2006, \apj, 650, 128

\bibitem[{{Krempec-Krygier} \& {Krygier}(1999)}]{1999AcA....49..403K}
{Krempec-Krygier}, J., \& {Krygier}, B. 1999, Acta Astronomica, 49, 403

\bibitem[{{Liedahl} {et~al.}(1995){Liedahl}, {Osterheld}, \&
  {Goldstein}}]{1995ApJ...438L.115L}
{Liedahl}, D.~A., {Osterheld}, A.~L., \& {Goldstein}, W.~H. 1995, \apjl, 438,
  L115

\bibitem[{{Markevitch} {et~al.}(1998){Markevitch}, {Forman}, {Sarazin}, \&
  {Vikhlinin}}]{1998ApJ...503...77M}
{Markevitch}, M., {Forman}, W.~R., {Sarazin}, C.~L., \& {Vikhlinin}, A. 1998,
  \apj, 503, 77

\bibitem[{{Markevitch} \& {Vikhlinin}(2007)}]{2007PhR...443....1M}
{Markevitch}, M., \& {Vikhlinin}, A. 2007, \physrep, 443, 1

\bibitem[{{Markevitch} {et~al.}(2001){Markevitch}, {Vikhlinin}, \&
  {Mazzotta}}]{2001ApJ...562L.153M}
{Markevitch}, M., {Vikhlinin}, A., \& {Mazzotta}, P. 2001, \apjl, 562, L153

\bibitem[{{Markevitch} {et~al.}(1996){Markevitch}, {Sarazin}, \&
  {Irwin}}]{1996ApJ...472L..17M}
{Markevitch}, M.~L., {Sarazin}, C.~L., \& {Irwin}, J.~A. 1996, \apjl, 472, L17+

\bibitem[{{Marshall} {et~al.}(2004){Marshall}, {Tennant}, {Grant}, {Hitchcock},
  {O'Dell}, \& {Plucinsky}}]{aciscontaminant}
{Marshall}, H.~L., {Tennant}, A., {Grant}, C.~E., {Hitchcock}, A.~P., {O'Dell},
  S.~L., \& {Plucinsky}, P.~P. 2004, in Presented at the Society of
  Photo-Optical Instrumentation Engineers (SPIE) Conference, Vol. 5165, X-Ray
  and Gamma-Ray Instrumentation for Astronomy XIII. Edited by Flanagan, Kathryn
  A.; Siegmund, Oswald H. W. Proceedings of the SPIE, Volume 5165, pp. 497-508
  (2004)., ed. K.~A. {Flanagan} \& O.~H.~W. {Siegmund}, 497--508

\bibitem[{{Martini} {et~al.}(2004){Martini}, {Kelson}, {Mulchaey}, \&
  {Athey}}]{2004cgpc.sympE..31M}
{Martini}, P., {Kelson}, D.~D., {Mulchaey}, J.~S., \& {Athey}, A. 2004, in
  Clusters of Galaxies: Probes of Cosmological Structure and Galaxy Evolution,
  ed. J.~S. {Mulchaey}, A.~{Dressler}, \& A.~{Oemler}

\bibitem[{{Mathiesen} \& {Evrard}(2001)}]{2001ApJ...546..100M}
{Mathiesen}, B.~F., \& {Evrard}, A.~E. 2001, \apj, 546, 100 (ME01)

\bibitem[{{Maughan}(2007)}]{2007astro.ph..3504M}
{Maughan}, B.~J. 2007, ArXiv Astrophysics e-prints

\bibitem[{{Maughan} {et~al.}(2007){Maughan}, {Jones}, {Forman}, \& {Van
  Speybroeck}}]{2007astro.ph..3156M}
{Maughan}, B.~J., {Jones}, C., {Forman}, W., \& {Van Speybroeck}, L. 2007,
  ArXiv Astrophysics e-prints

\bibitem[{{Mazzotta} {et~al.}(2001{\natexlab{a}}){Mazzotta}, {Markevitch},
  {Forman}, {Jones}, {Vikhlinin}, \& {VanSpeybroeck}}]{2001astro.ph..8476M}
{Mazzotta}, P., {Markevitch}, M., {Forman}, W.~R., {Jones}, C., {Vikhlinin},
  A., \& {VanSpeybroeck}, L. 2001{\natexlab{a}}, ArXiv Astrophysics e-prints

\bibitem[{{Mazzotta} {et~al.}(2001{\natexlab{b}}){Mazzotta}, {Markevitch},
  {Vikhlinin}, {Forman}, {David}, \& {VanSpeybroeck}}]{2001ApJ...555..205M}
{Mazzotta}, P., {Markevitch}, M., {Vikhlinin}, A., {Forman}, W.~R., {David},
  L.~P., \& {VanSpeybroeck}, L. 2001{\natexlab{b}}, \apj, 555, 205

\bibitem[{{Mazzotta} {et~al.}(2004){Mazzotta}, {Rasia}, {Moscardini}, \&
  {Tormen}}]{2004MNRAS.354...10M}
{Mazzotta}, P., {Rasia}, E., {Moscardini}, L., \& {Tormen}, G. 2004, \mnras,
  354, 10

\bibitem[{{McCarthy} {et~al.}(2004){McCarthy}, {Balogh}, {Babul},
  {Poole}, \& {Horner}}]{2004ApJ...613..811M}
{{McCarthy}, I.~G. and {Balogh}, M.~L. and {Babul}, A. and {Poole},
  G.~B. \& {Horner}, D.~J.}, 2004, \apj, 613, 811

\bibitem[{{Mercurio} {et~al.}(2003){Mercurio}, {Massarotti}, {Merluzzi},
  {Girardi}, {La Barbera}, \& {Busarello}}]{2003A&A...408...57M}
{Mercurio}, A., {Massarotti}, M., {Merluzzi}, P., {Girardi}, M., {La Barbera},
  F., \& {Busarello}, G. 2003, \aap, 408, 57

\bibitem[{{Metzger} \& {Ma}(2000)}]{2000AJ....120.2879M}
{Metzger}, M.~R., \& {Ma}, C.-P. 2000, \aj, 120, 2879

\bibitem[{{Mewe} {et~al.}(1985){Mewe}, {Gronenschild}, \& {van den
  Oord}}]{1985A&AS...62..197M}
{Mewe}, R., {Gronenschild}, E.~H.~B.~M., \& {van den Oord}, G.~H.~J. 1985,
  \aaps, 62, 197

\bibitem[{{Mewe} {et~al.}(1986){Mewe}, {Lemen}, \& {van den
  Oord}}]{1986A&AS...65..511M}
{Mewe}, R., {Lemen}, J.~R., \& {van den Oord}, G.~H.~J. 1986, \aaps, 65, 511

\bibitem[{{Mohr} \& {Evrard}(1997)}]{1997ApJ...491...38M}
{Mohr}, J.~J., \& {Evrard}, A.~E. 1997, \apj, 491, 38

\bibitem[{{Molendi} {et~al.}(2000){Molendi}, {De Grandi}, \&
  {Fusco-Femiano}}]{2000ApJ...534L..43M}
{Molendi}, S., {De Grandi}, S., \& {Fusco-Femiano}, R. 2000, \apjl, 534, L43

\bibitem[{{Morrison} \& {McCammon}(1983)}]{1983ApJ...270..119M}
{Morrison}, R., \& {McCammon}, D. 1983, \apj, 270, 119

\bibitem[{{Nagai} {et~al.}(2007){Nagai}, {Kravtsov}, \&
  {Vikhlinin}}]{2007ApJ...668....1N}
{Nagai}, D., {Kravtsov}, A.~V., \& {Vikhlinin}, A., 2007, \apj, 668,
  1

\bibitem[{{Nousek} \& {Shue}(1989)}]{1989ApJ...342.1207N}
{Nousek}, J.~A., \& {Shue}, D.~R. 1989, \apj, 342, 1207

\bibitem[{{Poole} {et~al.}(2006){Poole}, {Fardal}, {Babul},
  {McCarthy}, {Quinn}, \& {Wadsley}}]{2006MNRAS.373..881P}
{Poole}, G.~B., {Fardal}, M.~A., {Babul}, A., {McCarthy}, I.~G.,
  {Quinn}, T., \& {Wadsley}, J. 2006, \mnras, 373, 881

\bibitem[{{O'Hara} {et~al.}(2006){O'Hara}, {Mohr}, {Bialek}, \&
  {Evrard}}]{2006ApJ...639...64O}
{O'Hara}, T.~B., {Mohr}, J.~J., {Bialek}, J.~J., \& {Evrard}, A.~E. 2006, \apj,
  639, 64

\bibitem[{{Ohta} {et~al.}(2001){Ohta}, {Kumai}, {Watanabe}, {Furuzawa},
  {Akimoto}, {Tawara}, {Sato}, {Yamashita}, {Arai}, {Shiratori}, {Miyoshi}, \&
  {Mazure}}]{2001ASPC..251..474O}
{Ohta}, Y., {Kumai}, Y., {Watanabe}, M., {Furuzawa}, A., {Akimoto}, F.,
  {Tawara}, Y., {Sato}, S., {Yamashita}, K., {Arai}, K., {Shiratori}, Y.,
  {Miyoshi}, S., \& {Mazure}, A. 2001, in Astronomical Society of the Pacific
  Conference Series, Vol. 251, New Century of X-ray Astronomy, ed. H.~{Inoue}
  \& H.~{Kunieda}, 474--+

\bibitem[{{Ota} {et~al.}(2006){Ota}, {Kitayama}, {Masai}, \&
  {Mitsuda}}]{2006ApJ...640..673O}
{Ota}, N., {Kitayama}, T., {Masai}, K., \& {Mitsuda}, K. 2006, \apj, 640, 673

\bibitem[{{Peres} {et~al.}(1998){Peres}, {Fabian}, {Edge}, {Allen},
  {Johnstone}, \& {White}}]{1998MNRAS.298..416P}
{Peres}, C.~B., {Fabian}, A.~C., {Edge}, A.~C., {Allen}, S.~W., {Johnstone},
  R.~M., \& {White}, D.~A. 1998, \mnras, 298, 416

\bibitem[{{Riess} {et~al.}(1998){Riess}, {Filippenko}, {Challis},
  {Clocchiatti}, {Diercks}, {Garnavich}, {Gilliland}, {Hogan}, {Jha},
  {Kirshner}, {Leibundgut}, {Phillips}, {Reiss}, {Schmidt}, {Schommer},
  {Smith}, {Spyromilio}, {Stubbs}, {Suntzeff}, \&
  {Tonry}}]{1998AJ....116.1009R}
{Riess}, A.~G., {Filippenko}, A.~V., {Challis}, P., {Clocchiatti}, A.,
  {Diercks}, A., {Garnavich}, P.~M., {Gilliland}, R.~L., {Hogan}, C.~J., {Jha},
  S., {Kirshner}, R.~P., {Leibundgut}, B., {Phillips}, M.~M., {Reiss}, D.,
  {Schmidt}, B.~P., {Schommer}, R.~A., {Smith}, R.~C., {Spyromilio}, J.,
  {Stubbs}, C., {Suntzeff}, N.~B., \& {Tonry}, J. 1998, \aj, 116, 1009

\bibitem[{{Riess} {et~al.}(2007){Riess}, {Strolger}, {Casertano}, {Ferguson},
  {Mobasher}, {Gold}, {Challis}, {Filippenko}, {Jha}, {Li}, {Tonry}, {Foley},
  {Kirshner}, {Dickinson}, {MacDonald}, {Eisenstein}, {Livio}, {Younger}, {Xu},
  {Dahl{\'e}n}, \& {Stern}}]{2007ApJ...659...98R}
{Riess}, A.~G., {Strolger}, L.-G., {Casertano}, S., {Ferguson}, H.~C.,
  {Mobasher}, B., {Gold}, B., {Challis}, P.~J., {Filippenko}, A.~V., {Jha}, S.,
  {Li}, W., {Tonry}, J., {Foley}, R., {Kirshner}, R.~P., {Dickinson}, M.,
  {MacDonald}, E., {Eisenstein}, D., {Livio}, M., {Younger}, J., {Xu}, C.,
  {Dahl{\'e}n}, T., \& {Stern}, D. 2007, \apj, 659, 98

\bibitem[{{Rosati} {et~al.}(1995){Rosati}, {della Ceca}, {Burg}, {Norman}, \&
  {Giacconi}}]{1995ApJ...445L..11R}
{Rosati}, P., {della Ceca}, R., {Burg}, R., {Norman}, C., \& {Giacconi}, R.
  1995, \apjl, 445, L11

\bibitem[{{Sakelliou} \& {Ponman}(2004)}]{2004MNRAS.351.1439S}
{Sakelliou}, I., \& {Ponman}, T.~J. 2004, \mnras, 351, 1439

\bibitem[{{Sakelliou} \& {Merrifield}(2000)}]{2000MNRAS.311..649S}
{Sakelliou}, I., \& {Merrifield}, M.~R. 2000, \mnras, 311, 649

\bibitem[{{Sanderson} {et~al.}(2006){Sanderson}, {Ponman}, \&
  {O'Sullivan}}]{2006MNRAS.tmp.1068S}
{Sanderson}, A.~J.~R., {Ponman}, T.~J., \& {O'Sullivan}, E. 2006, \mnras, 1068

\bibitem[{{Smith} {et~al.}(2005){Smith}, {Kneib}, {Smail}, {Mazzotta},
  {Ebeling}, \& {Czoske}}]{2005MNRAS.359..417S}
{Smith}, G.~P., {Kneib}, J.-P., {Smail}, I., {Mazzotta}, P., {Ebeling}, H., \&
  {Czoske}, O. 2005, \mnras, 359, 417

\bibitem[{{Snowden} {et~al.}(2007){Snowden}, {Mushotzky}, {Kuntz}, \&
  {Davis}}]{chanxmmdis}
{Snowden}, S.~L., {Mushotzky}, R.~M., {Kuntz}, K.~D., \& {Davis}, D.~S. 2007,
  ArXiv e-prints, 710

\bibitem[{{Teague} {et~al.}(1990){Teague}, {Carter}, \&
  {Gray}}]{1990ApJS...72..715T}
{Teague}, P.~F., {Carter}, D., \& {Gray}, P.~M. 1990, \apjs, 72, 715

\bibitem[{{Townsley} {et~al.}(2000){Townsley}, {Broos}, {Garmire}, \&
  {Nousek}}]{2000ApJ...534L.139T}
{Townsley}, L.~K., {Broos}, P.~S., {Garmire}, G.~P., \& {Nousek}, J.~A. 2000,
  \apjl, 534, L139

\bibitem[{{Tucker} {et~al.}(1998){Tucker}, {Blanco}, {Rappoport}, {David},
  {Fabricant}, {Falco}, {Forman}, {Dressler}, \&
  {Ramella}}]{1998ApJ...496L...5T}
{Tucker}, W., {Blanco}, P., {Rappoport}, S., {David}, L., {Fabricant}, D.,
  {Falco}, E.~E., {Forman}, W., {Dressler}, A., \& {Ramella}, M. 1998, \apjl,
  496, L5+

\bibitem[{{Ventimiglia} {et~al.}(2008){Ventimiglia}, {Voit}, {Borgani}, \&
  {Donahue}}]{VV08}
{Ventimiglia}, D., {Voit}, G.~M., {Borgani}, S., \& {Donahue}, M. 2008, ApJ
  Submitted

\bibitem[{{Vikhlinin}(2006)}]{2006ApJ...640..710V}
{Vikhlinin}, A. 2006, \apj, 640, 710

\bibitem[{{Vikhlinin} {et~al.}(2005){Vikhlinin}, {Markevitch}, {Murray},
  {Jones}, {Forman}, \& {Van Speybroeck}}]{2005ApJ...628..655V}
{Vikhlinin}, A., {Markevitch}, M., {Murray}, S.~S., {Jones}, C., {Forman}, W.,
  \& {Van Speybroeck}, L. 2005, \apj, 628, 655

\bibitem[{{Vikhlinin} {et~al.}(1998){Vikhlinin}, {McNamara}, {Forman}, {Jones},
  {Quintana}, \& {Hornstrup}}]{1998ApJ...502..558V}
{Vikhlinin}, A., {McNamara}, B.~R., {Forman}, W., {Jones}, C., {Quintana}, H.,
  \& {Hornstrup}, A. 1998, \apj, 502, 558

\bibitem[{{Voit}(2005)}]{2005RvMP...77..207V}
{Voit}, G.~M. 2005, Reviews of Modern Physics, 77, 207

\bibitem[{{Wang} \& {Steinhardt}(1998)}]{1998ApJ...508..483W}
{Wang}, L., \& {Steinhardt}, P.~J. 1998, \apj, 508, 483

\bibitem[{{Wang} {et~al.}(2004){Wang}, {Khoury}, {Haiman}, \&
  {May}}]{2004PhRvD..70l3008W}
{Wang}, S., {Khoury}, J., {Haiman}, Z., \& {May}, M. 2004, \prd, 70, 123008

\bibitem[{{White} {et~al.}(1997){White}, {Jones}, \&
  {Forman}}]{1997MNRAS.292..419W}
{White}, D.~A., {Jones}, C., \& {Forman}, W. 1997, \mnras, 292, 419

\bibitem[{{Yang} {et~al.}(2004){Yang}, {Huo}, {Zhou}, {Xue}, {Mao}, {Ma}, \&
  {Chen}}]{2004ApJ...614..692Y}
{Yang}, Y., {Huo}, Z., {Zhou}, X., {Xue}, S., {Mao}, S., {Ma}, J., \& {Chen},
  J. 2004, \apj, 614, 692

\bibitem[{{Yuan} {et~al.}(2005){Yuan}, {Yan}, {Yang}, \&
  {Zhou}}]{2005ChJAA...5..126Y}
{Yuan}, Q.-R., {Yan}, P.-F., {Yang}, Y.-B., \& {Zhou}, X. 2005, Chinese Journal
  of Astronomy and Astrophysics, 5, 126

\end{thebibliography}
\end{document}